\RequirePackage{filecontents}

\documentclass[a4paper,fleqn,usenatbib,useAMS]{mnras}

\usepackage{graphicx}	% Including figure files
\usepackage{amsmath}	% Advanced maths commands
\usepackage{amssymb}	% Extra maths symbols
\usepackage{longtable}
\newcommand{\dd}{{\rm d}}
\newcommand{\comovingdist}{D_{\rm C}}

\newcommand{\vcmb}{v_{\rm cmb}}
\newcommand{\zcmb}{z_{\rm cmb}}

\usepackage{multicol}        % Multi-column entries in tables
\usepackage{bm}		% Bold maths symbols, including upright Greek
\usepackage{pdflscape}	% Landscape pages
\usepackage[table,xcdraw]{xcolor}

\usepackage[T1]{fontenc}
\usepackage{ae,aecompl}

\usepackage{multirow}

\usepackage{txfonts}
\newcommand{\newthree}[1]{\textcolor{black}{#1}}
\title[The effects of $v_{pec}$ in SNIa environments on $H_{0}$]{The effects of peculiar velocities in SNIa environments \\on the local $H_{0}$ measurement}

\author[T. M. Sedgwick et al.]{Thomas M. Sedgwick\thanks{Contact e-mail: \href{mailto:T.M.Sedgwick@2013.ljmu.ac.uk}{T.M.Sedgwick@2013.ljmu.ac.uk}},
Chris A. Collins, Ivan K. Baldry \& Philip A. James
\\
Astrophysics Research Institute, Liverpool John Moores University, IC2, Liverpool Science Park, 146 Brownlow Hill, L3 5RF
}

\date{Accepted by MNRAS, 2020 November.}

\pubyear{2020}

\begin{document}
\label{firstpage}
\pagerange{\pageref{firstpage}--\pageref{lastpage}}
\maketitle

\begin{abstract}
The discrepancy between estimates of the Hubble Constant ($H_{0}$) measured from local {($z~\lesssim~0.1$)} scales and from scales of the sound horizon is a crucial problem in modern cosmology. Peculiar velocities {($v_{pec}$)} of standard candle distance indicators can systematically affect local $H_{0}$ measurements. We here use 2MRS galaxies to measure the local galaxy density field, finding a notable {$z~<~0.05$} underdensity in the SGC-6dFGS region of $27~\pm~2$~\%. However, no strong evidence for a `Local Void' pertaining to the full 2MRS sky coverage is found. Galaxy densities are used to measure a density parameter, $\Delta \phi_{+-}$, {which we introduce as a proxy} for {$v_{pec}$} which quantifies density gradients along a SN line-of-sight. $\Delta \phi_{+-}$ is found to correlate with local $H_{0}$ estimates from \newthree{88 Pantheon} SNeIa {($0.02~<~z~<~0.05$)}. Density structures on scales of $\sim 50$\,Mpc are found to correlate strongest with $H_{0}$ estimates in both the observational data and in mock data from the MDPL2-Galacticus simulation.
{Using trends of $H_{0}$ with $\Delta \phi_{+-}$,} we can correct for the effects of density structure on local $H_{0}$ estimates, even in the presence of biased {$v_{pec}$}.
{However, the difference in the inferred $H_0$ estimate with and without the peculiar velocity correction is limited to $<~0.1$ \%. We conclude that accounting for environmentally-induced peculiar velocities of SNIa host galaxies does not resolve the tension between local and CMB-derived $H_0$ estimates.}

\end{abstract}

\begingroup
\let\clearpage\relax
\endgroup
\newpage 

\section{Introduction}\label{sec:intro}

The Hubble Constant at the present-epoch ($H_{0}$) parameterises the current rate of expansion of the Universe. A knowledge of the precise value of $H_{0}$ is crucial to $\Lambda$-CDM simulations and their extensions, to our description of the present-day Universe and to predictions of its ultimate fate.

A key problem in modern day cosmology is the persistent tension between measurements of $H_{0}$ when probed on different scales. Using measurements of anisotropies in the cosmic microwave background (CMB) and calibrating using a $\Lambda$-CDM cosmology, the \citealt[][henceforth, P18]{P18}, obtain the most stringent estimate of $H_{0}$ from the physics of the sound horizon to date, finding {$H_{0}$~=~67.36~$\pm$~0.54~km~s$^{-1}$~Mpc$^{-1}$}. Alternatively, measurements of $H_{0}$ on local scales of our Universe find larger values of $H_{0}$ \citep{R16,R18}. \citet[][henceforth, R19]{R19}, using Large Magellanic Cloud (LMC) Cepheids to calibrate SNIa photometry, give an estimate of {$H_{0}$~=~74.03~$\pm$~1.42~km~s$^{-1}$~Mpc$^{-1}$}, a result which lies in {4.4$\sigma$} tension with that of P18.

Increasing numbers of works in the literature cite physical effects as the cause of the Hubble tension \citep[see, e.g.][]{DIV18,AGR19,VAT19}. Indeed, R19 note that the $H_{0}$ discrepancy may point towards a problem for $\Lambda$-CDM, given the reliance of sound-horizon-scale results on the assumption of the standard cosmology.

An alternative source of the Hubble tension could instead relate to local systematics: the cosmic distance ladder, utilised on local (typically, {$z \lesssim 0.1$}) scales (and for example in R19), offers a direct and largely model-independent measure of $H_{0}$. However, a problem faced on these scales is that peculiar velocities, due to the inhomogeneity of the local Universe, are non-negligible when compared to recession velocities. The component of an object's velocity due to cosmic expansion must be sufficiently decoupled from peculiar velocity for an accurate calculation of $H_{0}$. Peculiar velocities are, on local scales, solely gravitationally induced motions \citep{PEE82}, and as a result, these velocities are expected to be strongly correlated with the galaxy density field. 

There exists in the literature debated evidence for a `Local Void', or under-density at our location in the Universe. The contrast and isotropy of such an under-density {has been investigated} using various phenomena, including SNeIa \citep{ZEH98,JHA07,CON07}, clusters \citep{GIO99,HUD04,BOH15} and galaxies \citep{SHA84,HUA97,RAT96,BUS04,KEE13} to probe the local density. \citet[][henceforth, WS14]{WS14} find a particularly significant galaxy under-density, most prominent in the direction of the 6dFGS South Galactic Cap region (SGC-6dFGS) in which a deficit of $\sim$ 40~\% is estimated for {$z < 0.05$}. This region has been cited as under-dense independently from the galaxy samples of the 6dFGS Redshift Survey \citep{BUS04} and 2MASS \citep{FRI03}.

The above studies probe the density on a regional basis, and a stem of this debate is whether the local under-density found in numerous works would persist across the full sky \citep{S18,S19,R18b,R18c,KEN19}. Recent work from \citet{BOH19} finds a local X-ray cluster under-density which pertains to the full sky. The existence of such an isotropic void would be expected to induce a bias towards peculiar velocities away from the observer, typically increasing local $H_{0}$ estimates away from the true value. Whilst past studies have attempted to calculate the expected error in $H_{0}$ estimates from the measured density contrast \citep[see, e.g.][]{S19}, estimating the offset from the true $H_{0}$ often relies on a modelling of the void \citep{EM07,KEN19}.

In the present work, we first attempt to form an independent, near-full sky picture of the local galaxy density field for comparison with previous studies. We then introduce a method for the empirical estimation of peculiar velocities using the galaxy density field. To bypass assumptions related to the geometry of the Local Void, we instead directly search for correlations between the density field and SNIa $H_{0}$ estimates. In doing so, we demonstrate that peculiar velocities are more tightly linked to gradients in the density field along the SN LOS, than to the absolute density of the SN region. Ultimately, we are able to quantify the {fractional} effect of the galaxy density field on the local $H_{0}$ estimate.

The structure of the present work is as follows: Section~\ref{sec:Ivan} presents the Hubble constant estimator used in this study. Section~\ref{sec:Data} outlines the data sets used. Section~\ref{sec:rho} presents the methodology for the calculation of the local galaxy density field. Section~\ref{sec:SNIa} then discusses the application of the aforementioned $H_0$ prescription to a sample of SNeIa. We then introduce a density parameter using our galaxy density field, which is designed to act as a proxy for peculiar velocity. We test correlations of this parameter with our aforementioned SNIa $H_{0}$ estimates {in Section~\ref{sec:H0U21}}. In Section~\ref{sec:Sims} we repeat our analyses using mock data to compute a mock density field where line-of-sight velocities are known, in order to test our observational results and assess sources of uncertainty in the observations. {We conclude this section with final estimates of the fractional effect on the local $H_{0}$ measurement due to peculiar velocities.}

\subsection{Estimator for the locally-derived Hubble constant}\label{sec:Ivan}

{In this paper, the estimator for the measured Hubble constant is given by 
\begin{equation}\label{eq:ivanH0}
  H_{0,\rm est} 
  \: = \:  H_{0,\rm fid} \frac{D_{C,{\rm fid}}(\zcmb)}{D_{C,{\rm est}}}
  \: = \: \frac{c \int_0^{\zcmb} \, [E(z)]^{-1} \, \mathrm{d} z }{D_{C, \rm est}} \mbox{~~~,}
\end{equation}
where the terms with subscript `fid' correspond to the fiducial cosmology applied to 
calculate distances as a function of $\zcmb$, 
and $D_{C,{\rm est}}$ is the estimated comoving distance of the standard candle
[$D_C = D_L / (1+z_{\rm helio})$ assuming a flat cosmology].}
The CMB-frame redshift is given by 
\begin{equation}
1 + z_{\rm cmb} \: = \: (1 + z_{\rm helio}) (1 + z_{\rm sun,comp})
                \: = \: (1 + z_{\rm cos}) (1 + z_{\rm pec}) \mbox{~~~,}
\label{eqn:cmb-frame}
\end{equation}
where $z_{\rm sun,comp}$ is from the component of the Sun's motion toward the
source in the CMB frame, with $z_{\rm sun}=0.00123$ \citep{lineweaver96,Fixsen09},
and the other subscripts refer to the heliocentric, cosmological 
and peculiar redshifts of the observed source.

Defining velocity as $v = c \ln (1+z)$ (more useful and accurate than the historical $cz$, \citealt{BAL18}), 
a straightforward and transparent approximation for $D_C$, comoving distance, can be obtained 
using the usual {present-epoch} deceleration parameter ($q_0$) (see Appendix~\ref{sec:Ivancontd}). 
From Eqs.~\ref{eq:ivanH0} and~\ref{eqn:second-order-h-law}, 
an accurate approximation for the Hubble constant estimator is then given by 
\begin{equation}
  H_{0,\rm est} \: \simeq \: \frac{\vcmb}{D_{C,{\rm est}}} \left( 1 - \frac{q_{0, \rm fid} \vcmb}{2 c} \right) \mbox{~~~,}
\end{equation}
with $\vcmb = v_{\rm cos} + v_{\rm pec}$. 
From this equation, the effect of peculiar velocities and choice of fiducial 
cosmology on the estimated Hubble constant is evident. {The effects of cosmological assumptions on the results of the present work are shown in Section~\ref{sec:BAO}.}

Sources of uncertainty for estimating the Hubble constant include: 
(i) calibration of the standard candle scale, 
(ii) photometric measurements, 
(iii) bandshifts ($k$ corrections), 
(iv) evolution, 
(v) differences between the true cosmology and the fiducial cosmology, and
(vi) peculiar velocities.  
Any systematic uncertainty from the first two is generally independent of redshift, 
while the uncertainty from the cosmology (or bandshift or evolution) 
increases approximately proportional to $\vcmb$. 
The uncertainties from peculiar velocities are approximately proportional to $1/v_{\rm cos}$ because 
$\vcmb = v_{\rm cos} (1 + v_{\rm pec}/v_{\rm cos})$.

\begin{figure}
\centerline{\includegraphics[width=1.\columnwidth]{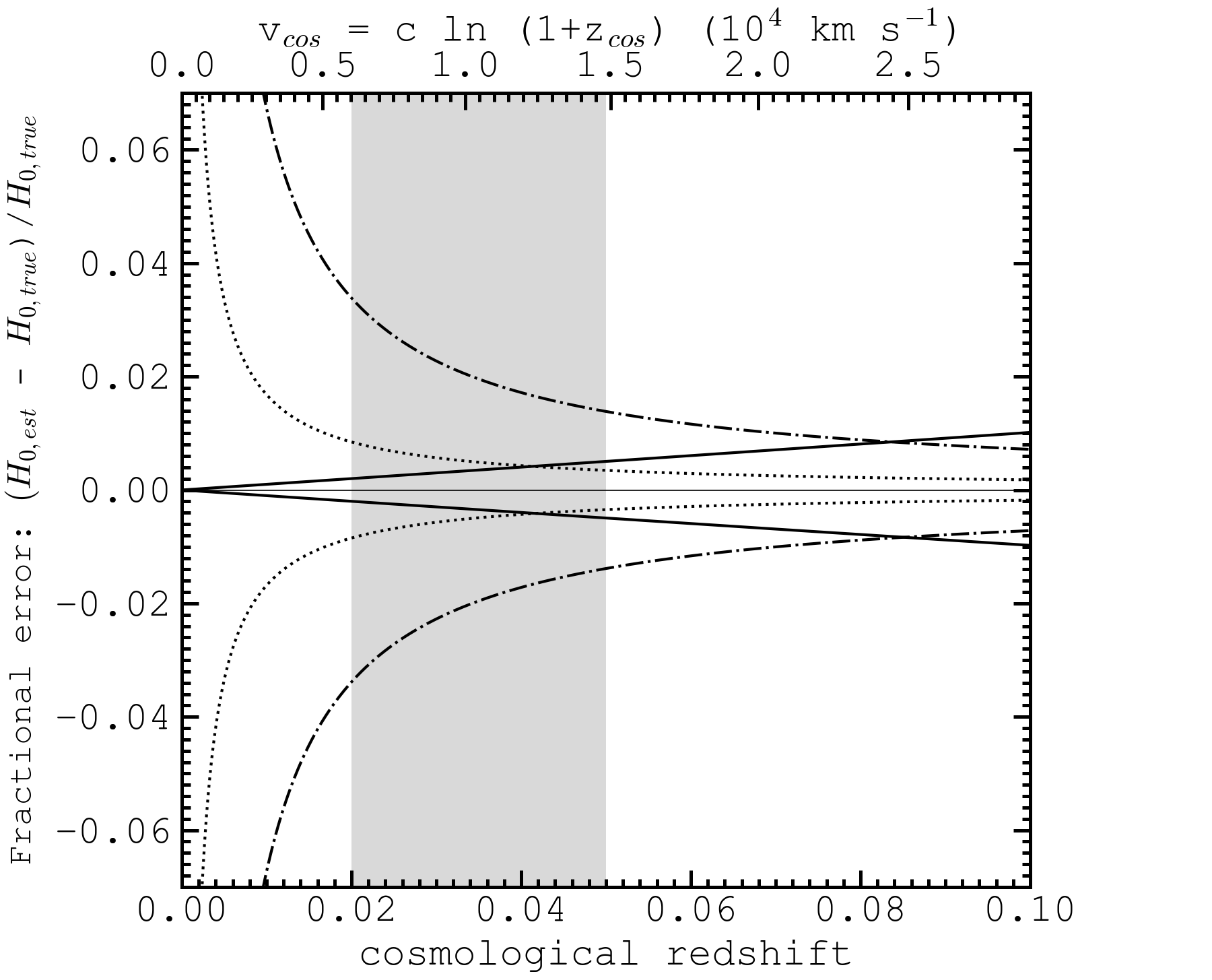}}
\caption{{The fractional error in the estimated Hubble constant due to peculiar velocities and erroneous cosmological assumptions.}
  The solid lines show the {fractional error} with a 0.2 offset in $q_0$ between the true and 
  fiducial cosmologies.
  The dot-dashed lines show the {fractional error} for systematic offsets of 200\,km/s between 
  CMB-frame velocities and cosmological recession velocities, while the dotted lines show the same 
  with a reduced systematic offset of 50\,km/s. The shaded region depicts {$0.02 < z < 0.05$} which is ultimately the focus for $H_{0}$ estimates in the present work.}
\label{fig:offset}
\end{figure}

Figure~\ref{fig:offset} illustrates the differences in the $H_{0}$ estimate arising from redshift-dependent uncertainties. The impact of peculiar velocities, in particular any non-zero average, 
pushes one to measure $H_{0}$ at $v_{cos} > 20000{\,\rm km/s}$. However, in order to limit
the degeneracy with $q_0$ and uncertainties that scale proportional to $v_{cos}$, it would be useful
to measure $H_{0}$ at lower recessional velocities. Either way, it is important 
to control for any systematic peculiar velocity offsets in the standard candle sample. 
It is the aim of this paper to test and account for peculiar velocity biases. 

\section{Data} \label{sec:Data}

In order to quantify the effects of the galaxy density field on SNIa peculiar velocities, and hence on local $H_{0}$ measurements, we use 3 key data sources:

\begin{enumerate}
\item \textbf{The 2MASS Redshift Survey:} our galaxy sample with which to measure the galaxy density field must have redshifts and cover a large solid angle on the sky, in order to minimise biases due to cosmic variance. As such we utilise the 2MASS Redshift Survey (2MRS) from \citet{HUC12}, built from the Extended Source Catalogue (XSC) of the 2-Micron All-Sky Survey (2MASS) \citep{SKR06}. The result is a galaxy redshift sample of 44,599 galaxies with {$m_{K} \le 11.75$ mag} (henceforth, the $K$-band magnitude refers to the extinction-corrected 2MASS isophotal Vega magnitude measured in an elliptical aperture defined at 20 mag/sq.arcsec) and with {$|b| \ge 5^{\circ}$} ($|b| \ge 8^{\circ}$ for {$330^{\circ} < l \le 30^{\circ}$}, i.e. towards the Galactic bulge), giving 97.6\% completeness within these limits \citep{HUC12}, i.e. away from the Zone of Avoidance (ZoA). This high completeness coupled with redshift information allows the construction of a 3-dimensional picture of the local galaxy density field.

{\item \textbf{The Pantheon SNIa Sample:} to test for correlations of the local galaxy density field with $H_{0}$ measurements from SNeIa, we make use of the Pantheon SNIa Sample \citep{SCO18}. This sample compiles photometry and spectroscopic redshifts for 1048 SNeIa. In the present work we ultimately utilise \newthree{88} SNeIa which overlap with the 2MRS footprint, \newthree{are at least 50 Mpc from the ZoA in 3-dimensional Cartesian space}, and occupy the redshift range {{0.02 < $z$ < 0.05}}; the range for which our galaxy density field is best constrained. This is in order to produce the most reliable $H_0$ estimates or {fractional $H_{0}$ error} when corrected for peculiar velocities, as discussed in Section \ref{sec:H0U21}.}

\item \textbf{The MDPL2-Galacticus Simulation:} to test for the effects of sample volume, sample size, {and cosmic variance} on the strength of correlations of SNIa $H_{0}$ estimates with the density field, we will repeat our analyses using the mock data products of MDPL2-Galacticus \citep{KNE18}, produced by running the Galacticus semi-analytical code \citep{BEN12} on the MultiDark Planck 2 (MDPL2) hydrodynamical simulation \citep{KLY16}. Details are described in \citet{STO19} and in the above works, but to summarise: the result is a 1 $h^{-3}$ Gpc$^{3}$ box containing 3840$^3$ Dark Matter particles, whose SDSS $ugriz$ luminosities are traced over cosmic time. In the present work, we make use of the {$z$~=~0} redshift snapshot, using corresponding $z$-band galaxy luminosities to impose a detection-limit on the galaxy sample, in order to construct mock galaxy density {fields}, used for comparison with the 2MRS $K$-band observational counterpart. {We will also use these simulations to test for the cosmic variance on our results, and to estimate how likely our observed local density structure is within the present-day Universe.}
\end{enumerate}

\section{Methodology} \label{sec:Methodology}

\subsection{Measuring the 2MRS Galaxy Density Field}\label{sec:rho}

As discussed in Section \ref{sec:intro}, we aim to quantify the effects of the galaxy density field on SNIa peculiar velocities, and hence, on the local {estimate} of $H_{0}$. We therefore proceed to construct the galaxy density field from the 2MRS Galaxy Catalogue. 

\begin{figure}
    \centerline{\includegraphics[width=1.\columnwidth]{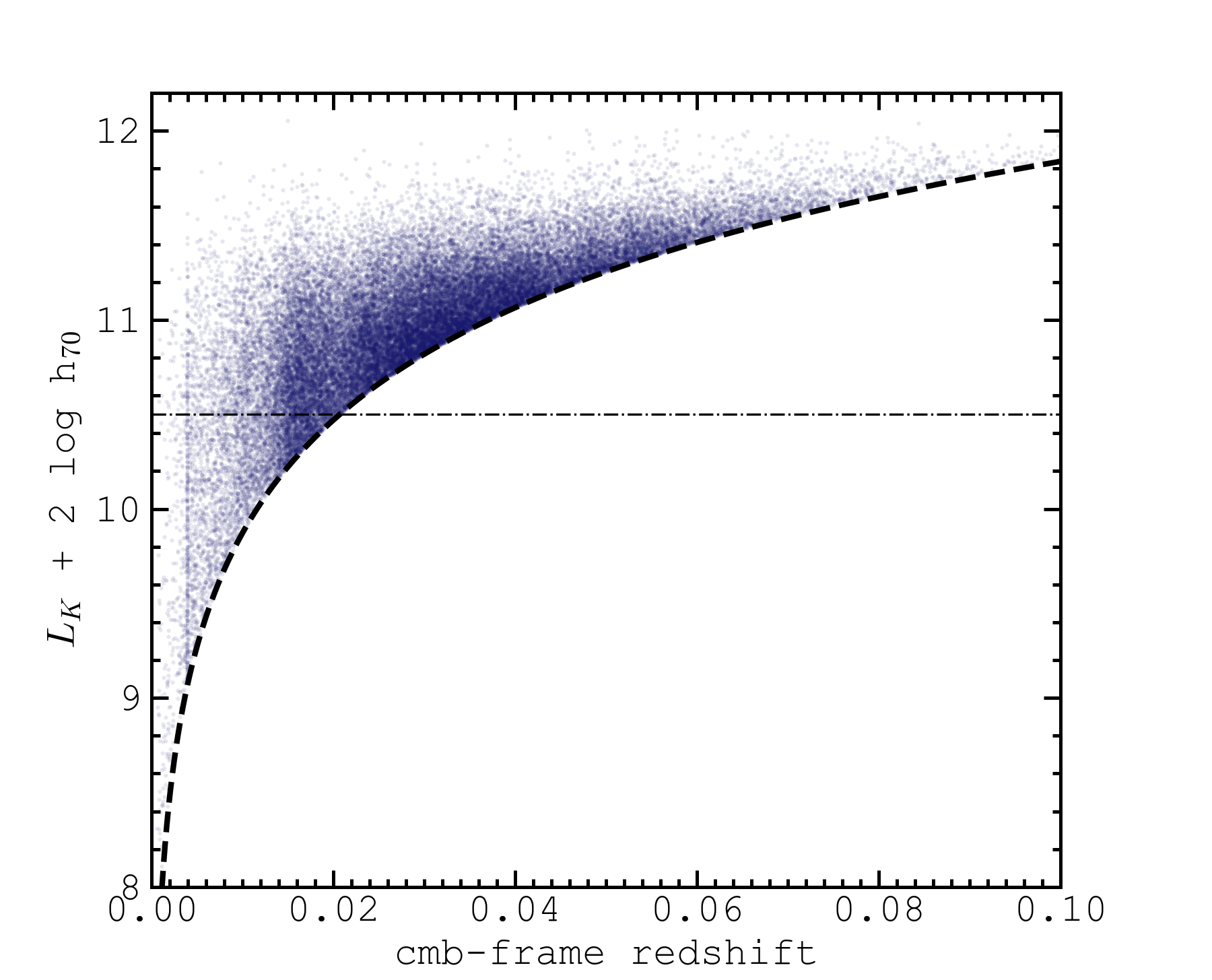}}
    \caption{2MRS galaxy $K$-band luminosities versus CMB-frame redshift. The dashed line indicates the flux limit as a function of redshift. Number densities as a function of redshift will be corrected to the number expected with {$L_{K} > 10.5$} (see text for details). {$L_{K} = 10.5$} is marked with the dot-dashed horizontal line.}
    \label{fig:1}
\end{figure}

This catalogue is flux-limited at {$m_{K} \le 11.75$}. As a result, we require a knowledge of the galaxy luminosity function from which to estimate the completeness of the sample as a function of redshift. Correcting for this completeness above a chosen luminosity value yields estimates of volume-limited number densities with redshift. We choose this minimum luminosity boundary to be {$L_{K} = 10.5$} (where $L_{K}$ here and henceforth refers to the luminosity in logarithmic units of the solar $K$-band luminosity quoted by \citealp{CWM03}). This gives volume-limited number densities for $z \lesssim 0.02$, and is chosen as a trade-off between the maximisation of statistics whilst limiting reliance on the completeness estimation method {which will be outlined}. The K-band luminosity distribution of the sample as a function of redshift is shown in Figure~\ref{fig:1}.

To improve the accuracy of the nearby galaxy density field, for which peculiar velocity is most troublesome for the determination of galaxy distance, we replace the 2MRS redshift in 2 cases: firstly, if the galaxy is matched within $5'$ {(on the sky)} and 150~km~s$^{-1}$ of a galaxy from the Updated Nearby Galaxy Catalogue of \citet{KAR13}, we utilise this catalogue distance. Secondly, if galaxies are matched within $0.5'$ of a member of the Extended Virgo Cluster Catalogue (EVCC) \citep{KIM14}, a distance of 16.5 Mpc is assumed. If either case applies, we compute and use the redshift implied from the comoving distance via a 737 cosmology ($H_{0}=70$, $\Omega _{\rm m}=0.3$, $\Omega _{\Lambda }=0.7$). Henceforth, the `fiducial cosmology' means 737 unless explicitly noted.

Galaxy $K$-band luminosities are calculated using Equation~\ref{eq:L}, where $M_{K,\odot}$ is the solar $K$-band Vega-mag absolute magnitude of 3.28, and $k(z)$ is the $k$-correction computed as {$k(z) = - 6.0 \log (1 + z_{hel})$} following \citet{KOC01}.

\begin{equation}\label{eq:L}
    L_{K} = \frac{5 \log (D_{L,fid}(\zcmb)/10 \, \rm pc) + M_{K,\odot} - m_{K} + k(z)}{2.5}
\end{equation}

To estimate the $K$-band luminosity function, we employ the parametric maximum-likelihood method of \citet[][henceforth, the STY method]{STY79}. The method is well-described in the literature, \citep[see, e.g.][]{LOV92}, but in short, we first assume that the galaxy luminosity distribution is well-described by a single-Schechter function \citep{SCH76}. We estimate the probability of observing a galaxy of a given luminosity at a given redshift. The single-set of Schechter function parameters $L^{*}$ (the `knee') and $\alpha$ (the faint-end slope), which maximises the product of these probabilities over the entire galaxy sample is our best maximum-likelihood estimate.

\begin{figure}
    \centerline{\includegraphics[width=1.\columnwidth]{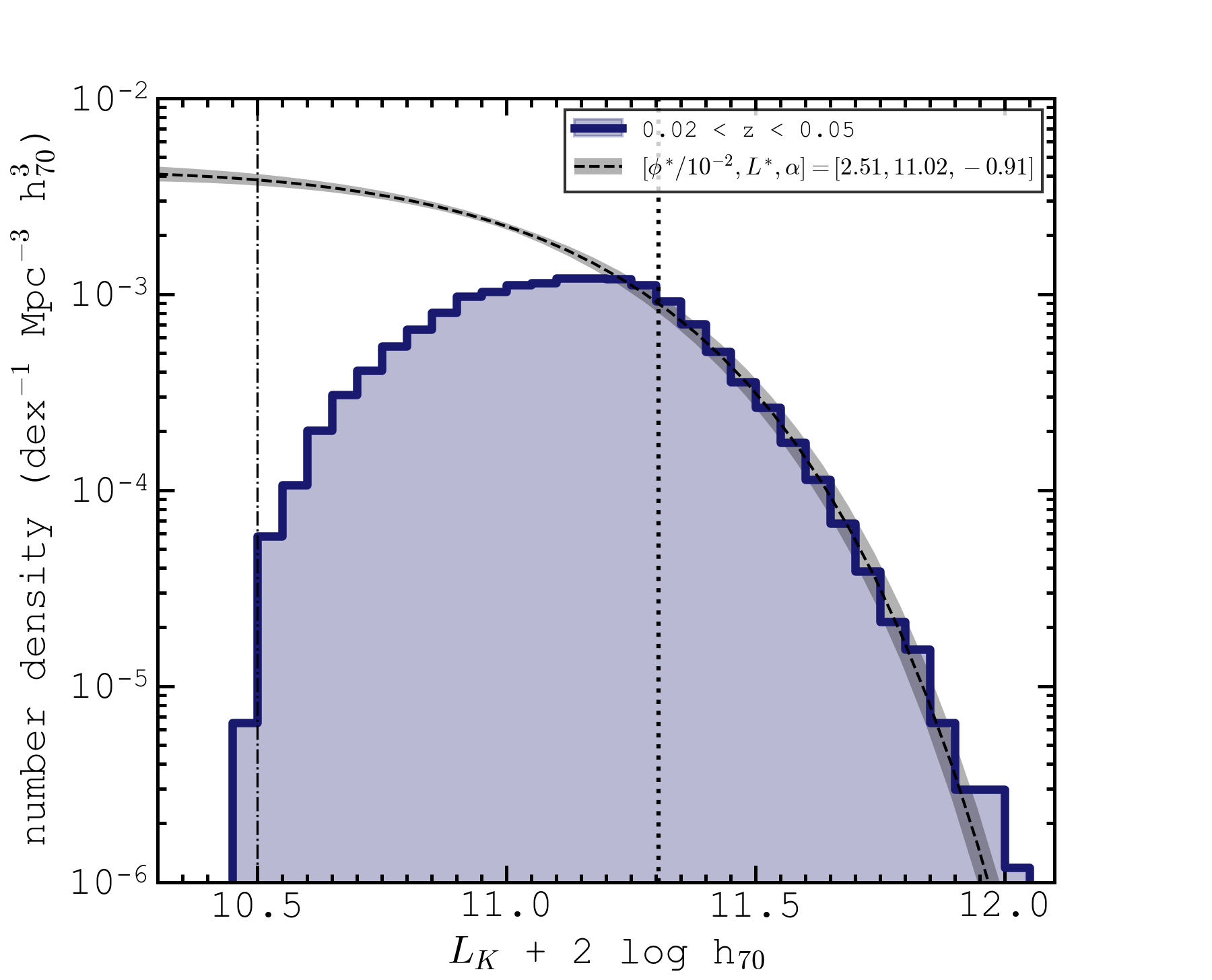}}
    \caption{{In blue: a histogram of the observed {0.02 < z < 0.05} $K$-band luminosity distribution of 2MRS galaxies. The dashed line shows the maximum-likelihood single-Schechter fit determined using the STY method. The dotted line shows the luminosity limit for {$z = 0.05$}. The dot-dashed line corresponds to {$L_{K} = 10.5$}.}}
    \label{fig:2}
\end{figure}

{The best-fit Schechter function is then used to estimate the completeness of galaxy number density at a given redshift. This is achieved by computing the ratio between the number density integrated above the flux limit corresponding to this redshift, and the integrated number density brighter than our reference luminosity of {$L_{K} = 10.5$}. For demonstrative purposes, Figure~\ref{fig:2} shows the luminosity distribution for the broad redshift range of {$0.02 < z < 0.05$}, as well as the maximum-likelihood Schechter function fit. For this redshift range, the STY method finds that the likelihood is maximised using parameters {$[L^{*},\alpha]=[11.02,-0.91]$}. Next, we assess the best-fit Schechter parameters in smaller redshift bins in order to quantify any parameter evolution.}

\begin{figure}
    \centerline{\includegraphics[width=1.\columnwidth]{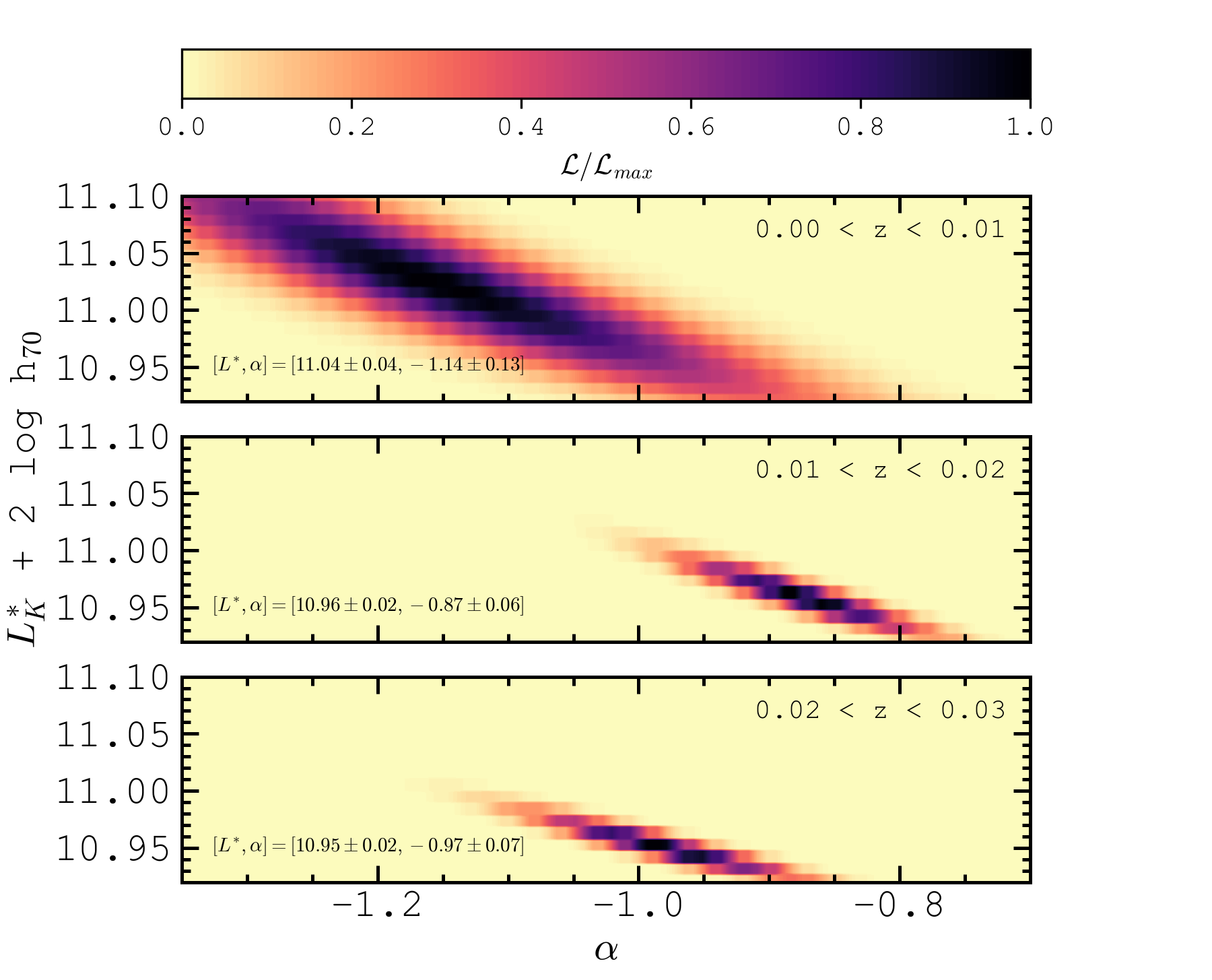}}
    \caption{Likelihood values for combinations of the single-Schechter function parameters $\alpha$ and $L^{*}$, from the STY method applied to the 2MRS $K$-band luminosities. Likelihood values are in units of the maximum likelihood in each panel. The three panels show 3 different CMB-frame redshift ranges of width 0.01, as indicated.}
    \label{fig:3}
\end{figure}

Figure~\ref{fig:3} shows the redshift evolution of these best-fit parameters. In redshift bins of width 0.01, likelihood values as a fraction of the maximum likelihood for each bin are assessed as a function of $L^{*}$ and $\alpha$. $1.8\sigma$ and $1.9\sigma$ separations in $L^{*}$ and $\alpha$, respectively, are found for {$0 < z < 0.01$} when compared with {$0.01 < z < 0.02$}. Comparing the latter bin with the {$0.02 < z < 0.03$} result, separations of $0.35\sigma$ and $1.08\sigma$ are found. We conclude that consistency is found within $2\sigma$ for the parameter values and hence adopt a fixed $\alpha$ value for the full redshift range. We use the value corresponding to the inverse-squared error weighted (henceforth, error-weighted) mean over all redshift bins out to {$z = 0.1$}, of $\alpha = -0.99$.

A correct assessment of luminosity versus redshift is crucial to analyses of the local density field. A lack of correction for this effect may result in an over-estimation of galaxy number densities which would worsen with increasing redshift. Such a slope to galaxy number density could lead to an over-estimate of the local outflow, which would lead to an under-estimation in local $H_{0}$ estimates.

\begin{figure}
    \centerline{\includegraphics[width=1.\columnwidth]{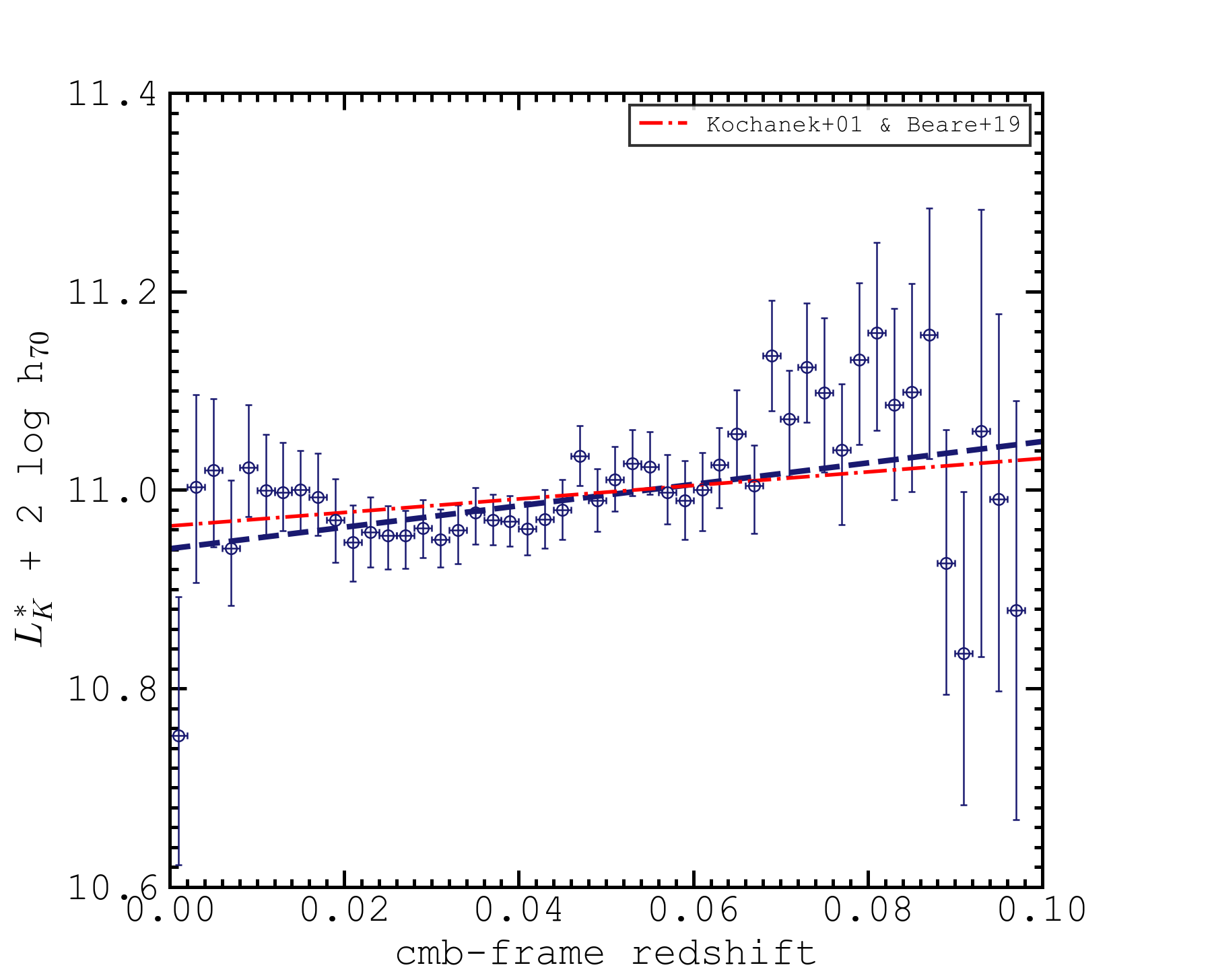}}
    \caption{The maximum-likelihood inferred values of the single-Schechter parameter $L^{*}$ as a function of CMB-frame redshift, when using a fixed {$\alpha = -0.99$} value. $L^{*}$ is computed in redshift bins of width 0.002. The error-weighted best-fit line is shown as the dashed blue line. The slope is consistent with that expected due to luminosity evolution, shown by the red dot-dashed line, made by connecting the $L^{*}$ values of \citet{KOC01} and \citet{BEA19} (see text).}
    \label{fig:4}
\end{figure}

{Galaxy luminosities may be expected to evolve since $z = 0.1$, primarily due to changes in mass-to-light ratio. The faint-end slope of the LF, $\alpha$, is not expected to evolve as significantly in this redshift range \citep[see, e.g.][]{MD14}. Irrespective of any $\alpha$ evolution, however, we can use the fact that $L^{*}$ and $\alpha$ are somewhat degenerate in order to treat any evolution as purely in luminosity, and as such this likely wraps-in changes to $\alpha$. (Furthermore, we find in {Section~\ref{subsec:regionalrho} that the choice of $\alpha$ does not affect results significantly.)} Repeating the Schechter fit determination as a function of redshift but with a fixed $\alpha$ value, we quantify the positive trend of $L^{*}$ with redshift, shown in Figure~\ref{fig:4}. The blue-dashed line shows the error-weighted regression fit, equating to {$L^{*} = 1.080 (z - 0.03) + 10.973$}, which has a Spearman rank correlation coefficient ($r_s$) of 0.558 and a $p$-value ($p$) of 0.001.}

An indication of expected luminosity evolution is shown as the red dashed line by connecting the inferred $K$-band $L^{*}$ value of \citet{KOC01} {($z < 0.01$)} with the {$z = 0.3$} value of \citet{BEA19} who adopt {$\alpha = -1.00$}. Our trend of $L^{*}$ with redshift is consistent with estimates of luminosity evolution found in the literature.

We next correct galaxy luminosities for evolution as a function of CMB-frame redshift, such that the evolution-corrected luminosity, $L'_{K}$, is given by {$L'_{K} = L_{K} + \delta_{L}$}, where {$\delta_{L} = - 1.080 (z - 0.03)$}. The sample is now re-selected with $L'_{K} > 10.5$.

\begin{figure}
    \centerline{\includegraphics[width=1.\columnwidth]{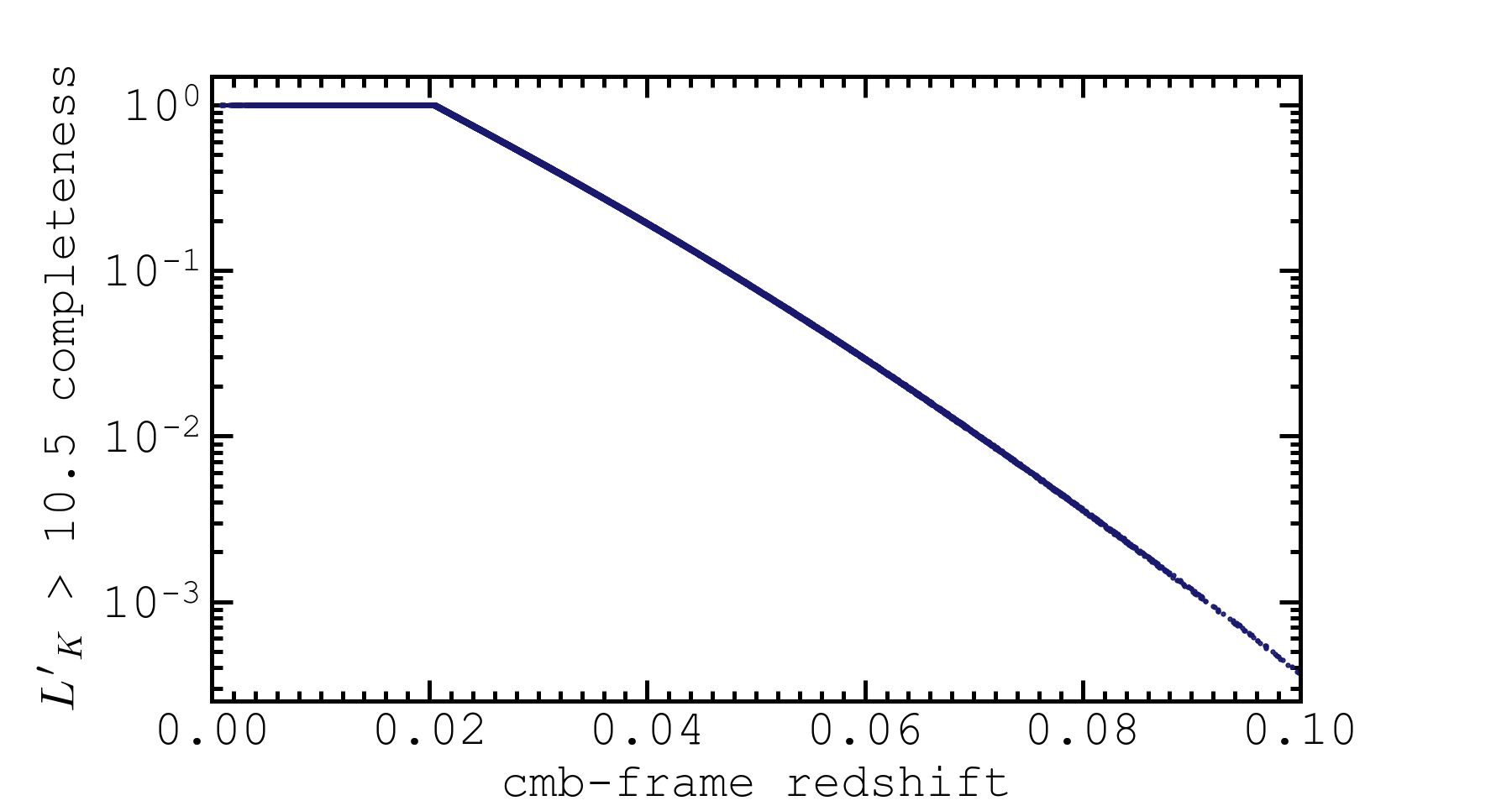}}
    \caption{Estimated completeness of {$L'_{K} > 10.5$} galaxy number statistics in the 2MRS galaxy sample, as a function of CMB-frame redshift.}\label{fig:completeness}
\end{figure}

With galaxy luminosities corrected for evolutionary effects, the luminosity function is well-approximated by the same single-Schechter function for the full redshift range {$(0 < z < 0.1)$}, with parameters {$[L^{*},\alpha] = [10.97,-0.99]$}. The sample completeness {above $L'_{K} = 10.5$} as a function of redshift, $\mathcal{C}(z)$, is estimated using Equation~\ref{eq:comp}, where $L'_{min}$ is the maximum of 10.5 and {$L_{K} + (m_K - 11.75)/2.5 + \delta_{L}$}. {Completeness as a function of redshift is shown in Figure~\ref{fig:completeness}. Galaxy counts are weighted by the inverse of $\mathcal{C}(z)$ where:}

\begin{equation}\label{eq:comp}
    \mathcal{C}(z) = \frac{\int_{L'_{min}}^{\infty}\phi(L')dL'}{\int_{10.5}^{\infty}  \phi(L')dL'}\mbox{~~~.}
\end{equation}

The volume-limited number density of galaxies in a redshift shell is estimated using Equation~\ref{eq:phi} where $\sum w_{N}$ is the sum of weights corresponding to galaxies within the shell, and $V$ is the shell volume, dependent on the solid angle spanned by the survey region:

\begin{equation}\label{eq:phi}
    \phi(z') = \frac{\sum w_{N}}{V}
\end{equation}

\subsection{A proxy for peculiar velocity from the galaxy density field}\label{sec:SNIa}

Equation~\ref{eq:ivanH0} shows that the local $H_{0}$ estimate inferred from a standard candle depends directly on the velocity of the object in the frame of the CMB. This velocity is the sum of components due to the expansion of the Universe (v$_{cos}$), and any peculiar velocities ($v_{pec}$). Hence, local $H_{0}$ estimates depend not only on cosmological expansion but also on $v_{pec}$ as demonstrated in Figure~\ref{fig:offset}.

In Section \ref{subsec:regionalrho} {we will} present galaxy number densities as a function of redshift, but for the {successive stages of our analysis we will require} a knowledge of the 3-dimensional galaxy density field. As mentioned in Section \ref{sec:intro}, the observed peculiar velocity is the line-of-sight (LOS) component of solely gravitationally induced motions on these scales. But it is not only the absolute density in a SNIa region that determines its peculiar velocity, but also the density gradient along the LOS \citep[see, e.g.][]{PEE80,LAH91}.

We require a density parameter which captures this LOS density gradient. This is achieved by measuring the density around the SN region in 2 hemispheres: {the density of galaxies in a hemisphere between the SN and observer is denoted $\phi_{-}$, and the density of galaxies in a hemisphere beyond the SN is denoted $\phi_{+}$.} The parameter $\Delta \phi_{+-}$ is then the {LOS density gradient in a SN environment, and can be written as:}

\begin{equation}\label{eq:delta_phi_pm}
    \Delta \phi_{+-} = \frac{\phi_{+} - \phi_{-}}{\phi_{+} + \phi_{-}} \mbox{~~~.}
\end{equation}

To determine the contributions of galaxies to $\Delta \phi_{+-}$, galaxy and SN positions are first converted into 3-dimensional Cartesian coordinates using RA, Dec, and comoving distance derived from CMB-frame redshift, using the fiducial cosmology. {We then measure the angle made between the LOS and the SN-galaxy directional vector. Let us define a function $\eta_{i}$. If the cosine of this angle is positive, $\eta_{i} = 1$, and a galaxy $i$ contributes to $\phi_{+}$. Otherwise $\eta_{i} = -1$ and the contribution is to $\phi_{-}$. $\Delta \phi_{+-}$ can now be re-written as:}

{\begin{equation}\label{eq:w}
    \Delta \phi_{+-} = \frac{\sum_{i}^{} \eta_{i} w_{N,i} \exp\left(-|\vec{r}_{gal,i}-\vec{r}_{sn}|^{2}/2 \sigma^{2})\right)}{\sum_{i}^{} w_{N,i} \exp\left(-|\vec{r}_{gal,i}-\vec{r}_{sn}|^{2}/2 \sigma^{2})\right)} \mbox{~~~.}
\end{equation}}

{Here, $w_{N,i}$ are the weights on contributions from each galaxy, $i$, determined previously with the STY method for our density vs redshift analysis. $\vec{r}_{gal,i}$ is the LOS vector from observer to each galaxy, and $\vec{r}_{sn}$ is the LOS vector from observer to SN. The parameter $\sigma$ controls how sharply contributions to $\Delta \phi_{+-}$ decrease as a Gaussian with SN-galaxy separation. We will refer to this parameter throughout the present work, along with another parameter, $R$, which represents the sphere radius out to which we consider density contributions.}

{We highlight the parameters $R$ and $\sigma$ because we aim to test for correlations for $H_{0}$ with $\Delta \phi_{+-}$. We will investigate whether particular values of $R$ and/or $\sigma$, maximise the strength of correlations, and in doing so, aim to reveal the scales of density structure which control peculiar velocities in SN environments.}

{Our method of estimating a proxy for peculiar velocity directly from the galaxy density field produces an independent test for the effects of density flows on $H_{0}$ estimates without the use of flow models, often utilised in the literature \citep[e.g.][]{HUD04,SCO18,NEI07}. We are able to assess the effects of peculiar velocity with no assumptions for the geometry of any density structure, and can assess the impact of structure on a wide variety of scales.}

\section{Results \& Discussion}
\subsection{Regional 2MRS Galaxy Densities}\label{subsec:regionalrho}

\begin{figure*}
    \includegraphics[width=0.93\textwidth]{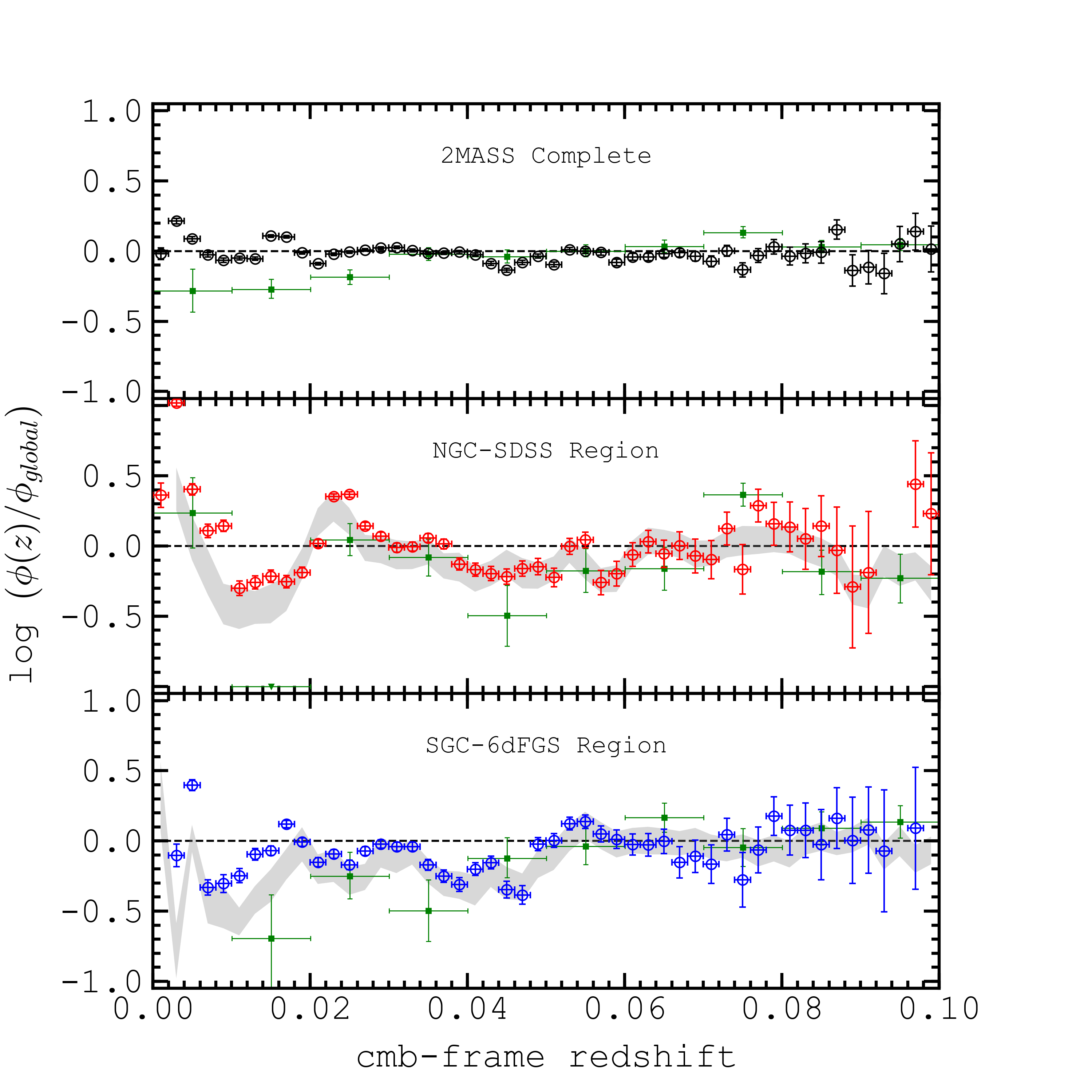}
    \caption{Galaxy number densities as a function of CMB-frame redshift in logarithmic units of the global density. Black, red and blue circles depict densities for the full 2MRS survey region, the NGC-SDSS region and SGC-6dFGS region, respectively. Shown as green points are $|b| > 20^{\circ}$ CLASSIX cluster densities (top), CLASSIX cluster densities in the NGC-SDSS region (middle) \citep{BOH19}, and REFLEX-II cluster densities in the SGC-6dFGS region. Grey-filled regions depict number densities found by WS14.}
    \label{fig:5}
\end{figure*}

The top panel of Figure~\ref{fig:5} shows galaxy number densities as a function of CMB-frame redshift for the sky coverage of 2MRS, equating to a $\sim 91\%$ coverage of the sky (see Section \ref{sec:Data}). Number densities are quoted in logarithmic units of the global density, $\phi_{global}$, itself calculated in the present work as the error-weighted {mean} density for {$0 < z < 0.1$}, with a value of {10$^{-2.49}$ Mpc$^{-3}$ bin$^{-1}$}. {Densities are given for redshift bins of width 0.002.} Poisson errors are shown, demonstrating the well-constrained nature of density structure out to at least {$z \sim 0.08$}. For the full 2MRS coverage, our {$z < 0.05$} integrated under-density equates to only $6 \pm 1 \%$. As such, {although we cannot make a strong statement for redshifts exceeding those of the 2MRS galaxy survey, we find no evidence for a void pertaining to the {full} sky out to at least $z = 0.1$}.

As a comparison with previous studies of the galaxy density field, we calculate densities for the regions of NGC-SDSS ({$150^{\circ}<$~RA~$<220^{\circ}$},~{$0^{\circ}<$~dec~$<50^{\circ}$}) and SGC-6dFGS ({$330^{\circ}<$~RA~$<50^{\circ}$},~{$-50^{\circ}<$~dec~$<0^{\circ}$}), regions of focus in WS14, who also utilise 2MASS photometry, coupled with redshifts from SDSS and 6dFGS for the 2 regions, respectively. Their densities are plotted as the grey filled regions in the bottom 2 panels of Figure~\ref{fig:5}, along with our results. Also plotted are the REFLEX-II/CLASSIX cluster densities from \citet{BOH15,BOH19}.

{A comparison with WS14 shows consistency for densities in the NGC-SDSS region}. We obtain an integrated {$z < 0.05$} {under}-density of $8 \pm 3 \%$ for this region. WS14 found their largest under-densities in the SGC-6dFGS region. {Calculating the integrated number density for {$z < 0.05$}, they obtain a $40 \pm 5 \%$ under-density in this region. We find an equivalent under-density in this region of} $27 \pm 2 \%$ (Poisson error only), which is a $2.4\sigma$ tension.

In light of this discrepancy we test our density measurements for the effects of our assumptions for the luminosity function, used to correct for $L_{K} > 10.5$ galaxy incompleteness beyond {$z \sim 0.02$}. We find that a deviation in the Schechter function slope of {$\alpha = 0.1$} either side of the adopted {$\alpha =-0.99$} produces a 3\% deviation to the {$z < 0.05$} integrated density, and as such cannot be the main source of the discrepancy. Note also that Figure~\ref{fig:5} shows our SGC-6dFGS result deviates most from the WS14 result for {$z < 0.02$}, the redshift range for which our sample is complete for {$L_{K} > 10.5$}, i.e. where no completeness corrections are required. {Furthermore, we estimate sample completeness using an evolving LF for $z\gtrsim0.02$, whereas WS14 use a fixed LF to model completeness for the full redshift range of {$0 < z < 0.1$}. {It is worth noting, however, that \citet{WS16} still find a significant local under-density, consistent with their previous analysis, when instead using a LF fitted simultaneously with the galaxy density distribution, albeit with a steeper faint-end slope to their LF than found in the present work.}}.

Comparing to other recent results in the literature, \citet{JL19} use physical Bayesian modelling of the non-linear matter distribution and find no clear evidence for an under-density in the direction of the SGC-6dFGS region, with an under-density of $3 \pm 11 \%$. \citet{BOH15} find a REFLEX-II cluster under-density in the SGC-6dFGS region of $55 \pm 10 \%$. Cluster bias is well-known to exaggerate voids and this is clear from Figure~\ref{fig:5}. Correcting for cluster bias they deduce a {$z < 0.05$} under-density comparable with that of the present work, of $20 \pm 8 \%$. {In Section~\ref{sec:SGCsims} we investigate the SGC-6dFGS under-density in more detail, using simulations to estimate how common such under-densities are in the Universe.}

{To summarise, we find no evidence for a significant void which pertains to the full sky, {out to the $z = 0.1$ limit of the 2MRS galaxy survey}. However, Figure~\ref{fig:5} shows that we reproduce well the regional density structures found by WS14, albeit with different amplitudes of the under-density of certain structures on scales of {$z < 0.05$}. Notable density structures reproduced in this work include the void in the direction of NGC-SDSS centred on $z \sim 0.015$ , for which we obtain a density $\sim 0.5\phi_{global}$, as well as the over-density on smaller scales ($z \sim 0.004$) in the same sky direction, of order 10 times that of the global density. Such density structures would be expected to be consequential for the peculiar velocities of SNeIa in these regions \citep[see, e.g.][]{PEE80,CBP81}. As such, quantifying and correcting for these effects is our main focus for the remainder of the present work.}

\subsection{{Correlations of $H_{0}$ with $\Delta \phi_{+-}$}}
\subsubsection{Pantheon SNe in the Galaxy Density Field}\label{sec:H0U21}

{We can estimate $H_0$ from individual SNeIa based on their redshifts and distance moduli (found by \citealt{SCO18}), using Equation~\ref{eq:ivanH0}. Note that this estimator is not sensitive to the fiducial value of $H_0$ assumed, and only slightly sensitive to differences in the assumption for $q_0$: as quantified in Section~\ref{sec:BAO}. However,  since our goal is to determine the effects of peculiar velocity, we choose to present the majority of results in terms of the \textit{fractional change} in $H_0$, which is not sensitive to the well-documented issue of SNIa distance calibration. The only exception is in Section~\ref{sec:BAO}, where for completeness, we give absolute $H_0$ estimates by calibrating SNIa distance moduli on the BAO-derived cosmic distance scale \citep{AND14}.}

{We calculate the fractional error in $H_0$ from the zero peculiar velocity case by performing an error-weighted linear fit of $\Delta \phi_{+-}$ to $H_0$. The fractional error in $H_0$ is then given as $(H_0 - c) / c$, where $c$ is the {$\Delta \phi_{+-}$ = 0} intercept of the regression line.} 
We use SNe with redshifts in the range {$0.02 < z < 0.05$} for this fit, as this range meets several useful criteria for our analyses: we see a trade off between uncertainties due to peculiar velocity and due to $q_{0}$ (See Figure \ref{fig:offset}); both the galaxy and SN statistics are high; the best-fit Schechter function parameters required to infer the density field are best-constrained; and it may be interesting to examine the effects of well-defined structures on peculiar velocities, found in this range (e.g. in NGC-SDSS and SGC-6dFGS). In short, this redshift range will produce the most reliable {estimates of fractional $H_{0}$ error due to peculiar velocity}. \newthree{111 of the 1048 Pantheon SNe are found in this range}.

For each SN, {if the nearest path to the edge of the 2MRS survey (i.e. to the ZoA) is shorter than $R$,} the SN is removed from the sample to prevent edge effects. {We also remove galaxies within 10~Mpc of the SN position. This is because the typical galaxy group velocity dispersion is a continuous scale from 10s of km~s$^{-1}$ (for groups of a few dwarf galaxies) to 1000s of km~s$^{-1}$ (for the richest clusters). Hence, the inferred line-of-sight group radius is of the order $\sim$10~Mpc for large groups. The positions of these galaxies relative to the SN are uncertain. Indeed, if included, these galaxies would also carry the most weight in our density prescription.}

\begin{figure*}
    \centerline{\includegraphics[width=0.96\textwidth]{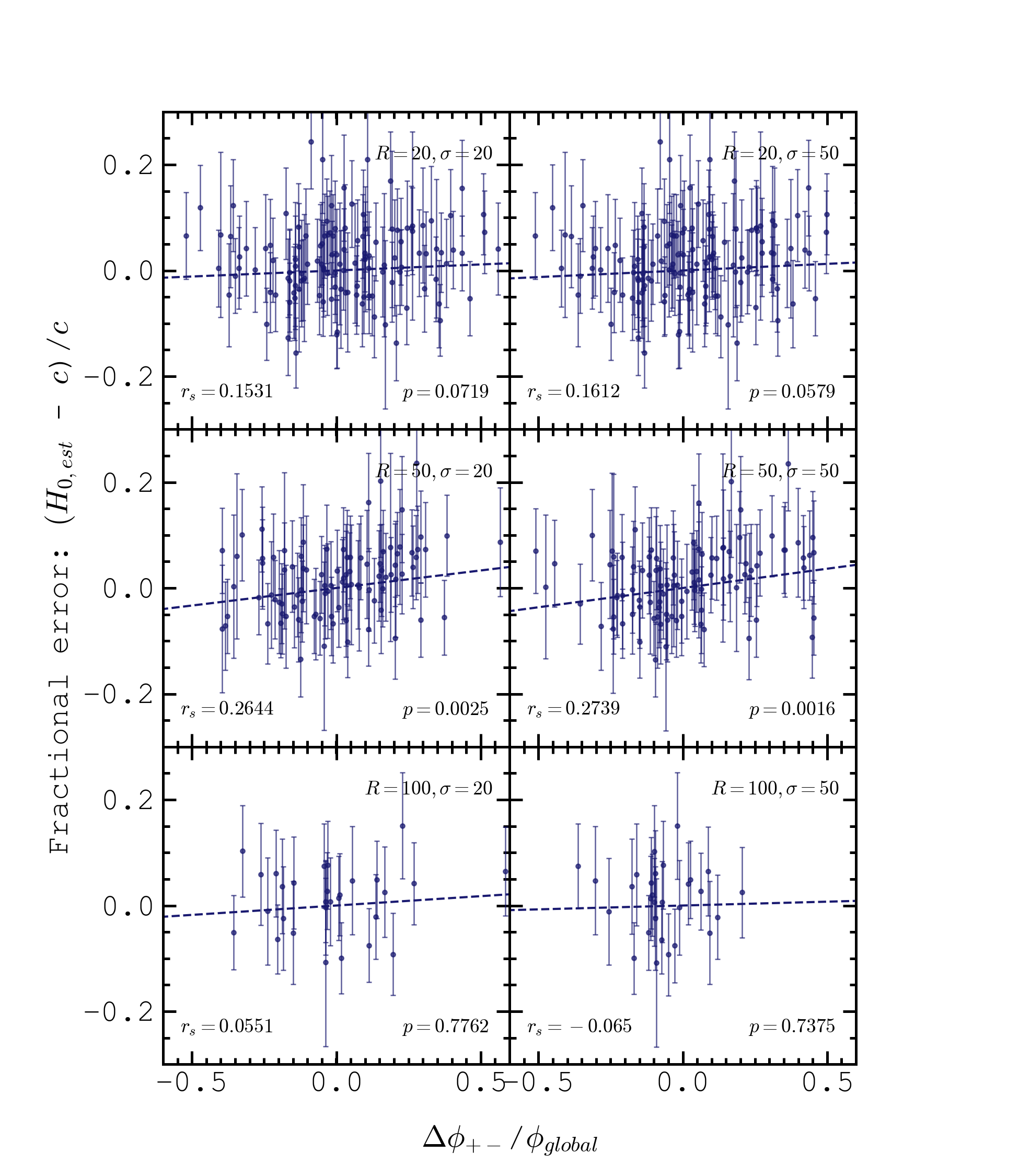}}
    \caption{{{Estimates of fractional error in $H_{0}$ for {0.02 < $z$ < 0.05} Pantheon SNe} as a function of {$\Delta \phi_{+-}$}. In each panel, {$\Delta \phi_{+-}$} is computed with different values for the Gaussian smoothing scale of density, around the SN ($\sigma$), and of the maximum separation from the SN considered in the density calculation ($R$). The error-weighted line-of-best fit to the data is shown for each $\sigma$-$R$ combination, as well as the Spearman's rank correlation coefficient ($r_{s}$) and $p$-value ($p$).}}
    \label{fig:7}
\end{figure*}

In Figure~\ref{fig:7}, the 6 panels show the differing strength of correlation of fractional $H_{0}$ error due to peculiar velocity with $\Delta \phi_{+-}$, as the sphere radius, $R$, and the density smoothing length, $\sigma$, are varied. In each panel, {the correlation is found to be roughly linear}, and so an error-weighted regression line is calculated. The corresponding {Spearman rank correlation coefficient ($r_{s}$) and $p$-value ($p$) are shown in each panel. We find that} the maximum significance of correlation between $H_{0}$ estimates and $\Delta \phi_{+-}$ (maximum $r_s$ and minimum $p$) arises for {[$R$,~$\sigma$]~=~[50~Mpc,~50~Mpc]}.

{The results shown adopt the cut within 10 Mpc of the SN, as discussed. This cut was found to reduce the $p$-value of the $H_{0}$ vs $\Delta \phi_{+-}$ fit by $\sim$~5\%. Using instead a 5~Mpc or 20~Mpc cut, we see in both cases a $\sim$~10\% rise to the $p$-value when compared to the preferred 10~Mpc cut.}

{{For the {$0.02 < z < 0.05$} Pantheon sample}, \newthree{88 out of 111} SNe are sufficiently far from the galaxy survey edge to assess the density within 50 Mpc of the SNe. For these \newthree{88} SNe we find {[$r_{s}$,~$p$]~=~[\newthree{0.2739,~0.0016}].}} Therefore, for the remainder of the present work, when referring to $\Delta \phi_{+-}$, we are using {[$R$,~$\sigma$]~=~[50~Mpc,~50~Mpc]} for its calculation. This result suggests that peculiar velocities are driven primarily by supercluster scale structure. In Section \ref{sec:Sims} we investigate and discuss this suggestion in more detail.

{We also investigate alternative prescriptions for our density parameter; We test for the change to correlations if galaxies within 10 Mpc of the SNe are instead included in the density measurements; We test correlations of the resultant density parameter with {fractional $H_{0}$ error} arising when using an inverse-squared weighting with separation. The observed peculiar velocity results from the net line-of-sight component of the gravitational force, and so an inverse squared weighting is expected to be most appropriate; We also test for the effects of modifying the density weights to also account for the luminosity of the galaxies, assuming that luminosity traces the galaxy mass. However, each of these prescriptions for $\Delta \phi_{+-}$ are found to correlate more weakly with {fractional $H_{0}$ error} than a Gaussian-smoothed number-density based calculation, albeit, marginally in the case of the 10 Mpc cut. For the remaining tests, this is likely due to the uncertainty in estimating the total (stellar + halo) galaxy mass from the luminosity. An over-weighting of individual galaxies can lead to a catastrophic miscalculation of the peculiar velocity proxy.}

\begin{figure*}
    \centerline{\includegraphics[width=1.2\textwidth]{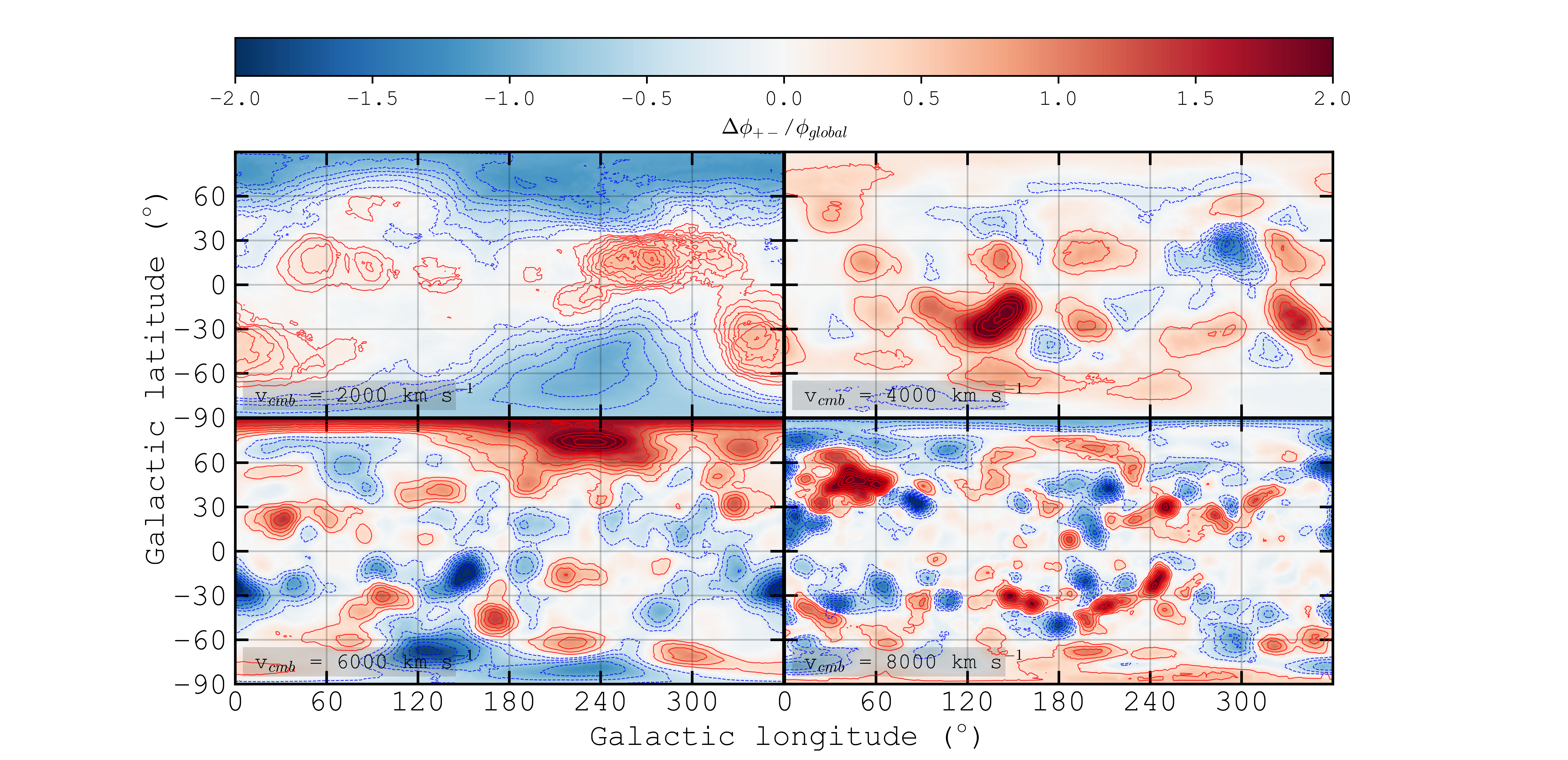}}
    \caption{The density parameter $\Delta \phi_{+-}$ (see text), plotted as a function of sky position, in Galactic coordinates. The parameter is assessed at 4 velocities in steps of {2000~km~s$^{-1}$}, where {$v~=~c~\ln~(1~+~z)$}, and approximately corresponding to distances from the observer. Objects in regions with {$\Delta \phi_{+-}$~>~0} are expected to flow away from the observer faster than the Hubble flow, and slower than the Hubble flow when {$\Delta \phi_{+-}$~<~0}.}
    \label{fig:6}
\end{figure*}

It was highlighted in Section \ref{sec:SNIa} that over- or under-density alone does not always result in significant peculiar velocities, and that galaxies at a density peak or trough, may experience a small \textit{net} force upon them and hence a small peculiar velocity. This is demonstrated using $\Delta \phi_{+-}$ (using {[$R$, $\sigma$] = [50 Mpc, 50 Mpc]}), in Figure~\ref{fig:6}, which shows the parameter as a function of sky position in Galactic co-ordinates. In each panel, the same process for calculating $\Delta \phi_{+-}$ around SNIa positions is applied to the whole sky, for different tomographic slices through the density field, at various steps of {$v_{cmb} = c \ln (1 + z_{cmb}$)}. 

Referring back to Figure~\ref{fig:5}, we saw a significant under-density centred on $z_{cmb}$~$\sim$~0.015, in the NGC-SDSS region. This redshift corresponds to a recession velocity in the CMB-frame of $\sim$~4000~km~s$^{-1}$. Note then, that in Figure~\ref{fig:6}, $\Delta \phi_{+-}$ is close to zero in the {$v_{cmb}$~=~4000~km~s$^{-1}$} panel. On the other-hand, at the redshifts corresponding to the {2000~km~s$^{-1}$} and {6000~km~s$^{-1}$} velocity slices, (0.007 and 0.020, respectively), objects are expected to be flowing away from the trough of under-density towards the over-dense peaks at {$z$ $\sim$ 0.003} and {$z$ $\sim$ 0.024}. This causes measurable effects on the values of $\Delta \phi_{+-}$ in the NGC-SDSS region, seen in Figure~\ref{fig:6}, with significant blueshift and redshift in the {$v_{cmb}$~=~2000} and {$v_{cmb}$~=~4000~km~s$^{-1}$} panels, respectively. This demonstrates how $\Delta \phi_{+-}$ is able to capture expected peculiar velocity information due to density gradients.

Another notable structural influence is the Perseus Cluster, situated at [$l$,~$b$,~$z_{cmb}$,~$v_{cmb}$~]~$\sim$~[150$^{\circ}$,~--13$^{\circ}$,~0.017,~5000~km~s$^{-1}$] \citep{PIF11}: in-fall to the cluster is seen to cause positive $\Delta \phi_{+-}$ (peculiar-velocity induced redshift) for the {$v_{cmb}$~=~4000~km~s$^{-1}$} slice, and negative $\Delta \phi_{+-}$ (peculiar-velocity induced blueshift) for the {$v_{cmb}$~=~6000~km~s$^{-1}$} slice.

\subsubsection{Mock data from MDPL2-Galacticus}\label{sec:Sims}

Note that in Figure~\ref{fig:7}, the mean value of $\Delta \phi_{+-}$ lies close to zero, implying that the {$0.02 < z < 0.05$} Pantheon SN sample is minimally biased in the sign of peculiar velocities. We also saw that $\Delta \phi_{+-}$ is correlated with locally-inferred fractional $H_{0}$ error estimates. We next turn to mock data from the MDPL2-Galacticus simulation \citep{KNE18} in order to test, firstly, whether trends of {fractional $H_{0}$ error} with $\Delta \phi_{+-}$ are consistent with the observations; secondly, whether the strength of correlation is limited by the ability of $\Delta \phi_{+-}$ to capture peculiar velocity information, or instead by observational photometric uncertainties, not present in the models; and finally, what the cosmic variance is in the $\Delta \phi_{+-}$ distribution, given our access to arbitrary observer positions. {This cosmic variance result gives us an estimate of the error on our observational peculiar velocity corrections.}

As discussed in Section \ref{sec:Data}, we utilise a {1 $h^{-3}$ Gpc$^{3}$} box with {3840$^3$} dark matter particles traced to the current epoch, using the {$z$~=~0} redshift snapshot. Each particle has 3-dimensional positions ($\vec{r}$) and velocities ($\vec{v}$). We then use the particles' mock $z$-band stellar luminosities to impose a detection-limit. $L_{z}$ is here defined as the logarithm of the luminosity in units of {4.4659e13 W Hz$^{-1}$}. The limit is then set to {$L_{z} = 8.843$} such that global `galaxy' density matches the global $L_{K} > 10.5$ density found for the 2MRS galaxy sample.

To calculate $\Delta \phi_{+-}$ and local {fractional $H_{0}$ errors} from the mock data, the observer's position in the 1 $h^{-3}$ Gpc$^{3}$ box is randomised, and the particle coordinates are redefined such that the observer lies at the origin. Next, galaxies lying at redshifts {$0.02 < z < 0.05$} from the observer, are selected at random as SNIa hosts. 

Peculiar velocities relative to the observer for all galaxies above the mock flux limit, including the SN hosts, are calculated as follows:
\begin{equation}\label{eq:vpecmod}
    v_{pec} = \frac{\vec{r}\cdot\vec{v}}{|\vec{r}|}  \mbox{~~~.}
\end{equation}
Galaxy redshifts due to cosmic expansion ($z_{cos}$) are inferred using the comoving distances {$D_{C} = |\vec{r}|$} associated via the fiducial cosmology. Mock observed redshifts ($z_{cmb}$) are then calculated using:
\begin{equation}\label{eq:zobsmod}
   \ln\;(1 + z_{cmb}) \: = \: \ln\;(1 + z_{cos}) + \frac{v_{pec}}{c}
   \mbox{~~~.}
\end{equation}

{Fractional $H_{0}$ errors} from the SNe are obtained using a modification of Equation~\ref{eq:ivanH0}:
\begin{equation}\label{eq:ivanH0mod}
   H_{0,\rm est} \: = \:
   H_{0,\rm fid} \frac{D_{C,{\rm fid}}(z_{cmb})}
                        {D_{C,{\rm fid}}(z_{cos})} \mbox{~~~.}
\end{equation}

\begin{figure}
    \centerline{\includegraphics[width=1.2\columnwidth]{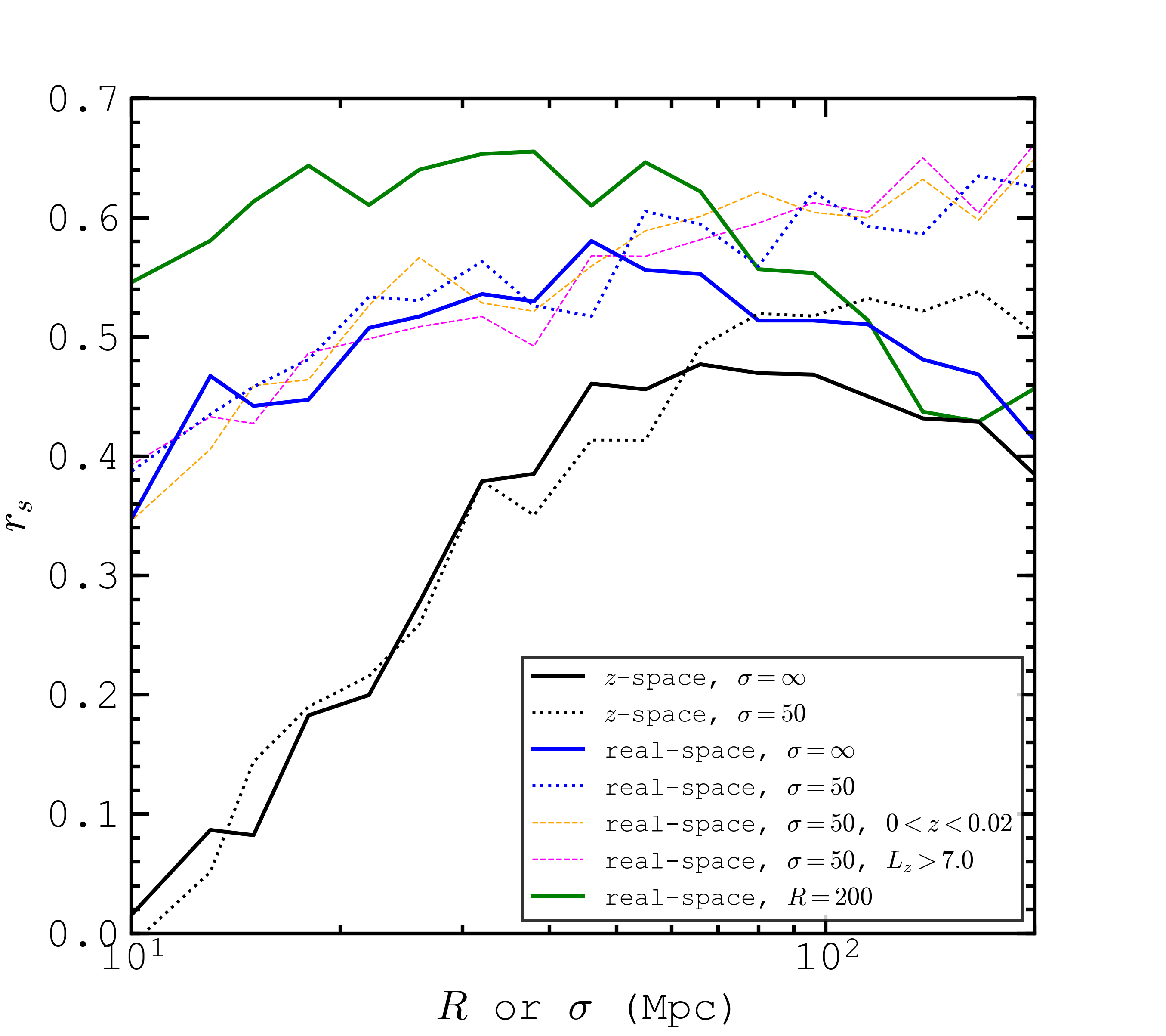}}
    \caption{Spearman rank correlation coefficients, $r_s$, corresponding to linear fits of {fractional $H_{0}$ error} to $\Delta \phi_{+-}$ for mock SNe positions in the {$z~=~0$} snapshot of the MDPL2 Galacticus simulation, as a function of $R$ or $\sigma$ in Mpc (used to calculate $\Delta \phi_{+-}$). $r_s$ is shown as a function of $R$ with the exception of the green solid line, where $r_s$ is shown as a function of $\sigma$. Unless stated, SNe are drawn from the simulation at redshifts {{$0.02 < z < 0.05$}} and the galaxy sample is luminosity limited at {$L_{z} > 8.843$} (see text for details).}
    \label{fig:phi_rs}
\end{figure}

$\Delta \phi_{+-}$ is finally calculated about the CMB-frame redshift-inferred SN positions, as {was} the observational data, using the resultant mock density field. 
Figure~\ref{fig:phi_rs} shows $r_s$ values corresponding to linear fits of $H_{0}$ to $\Delta \phi_{+-}$, where each fit is to 1000 random SN positions from the simulation. Values of {$10 < R < 200$ Mpc} are sampled, in equal logarithmic steps.

In black, mock-observed galaxy redshifts were used to produce the galaxy density field, to test for the effects of redshift space distortions on correlations. SNe with {$0.02 < z < 0.05$} were chosen to match the observations. $r_s$ is shown as a function of sphere size, $R$, within which $\Delta \phi_{+-}$ is calculated. The solid black line shows $\Delta \phi_{+-}$ when all galaxies contribute equally to the density. We observe that the maximum correlation of {fractional $H_{0}$ error} vs $\Delta \phi_{+-}$ comes for {$R \sim 50$ Mpc}. {When using a weighting of density contributions such that {$\sigma = 50$ Mpc}, we see that $r_{s}$ rises significantly as scales of {50 Mpc} are approached, and then improves marginally as this sphere size is increased further.} 

A benefit of the simulations is that we can repeat these tests but using the real-space positions of galaxies, as shown in blue. We observe once again a peak at {$R = 50$ Mpc} in the unweighted case, but the most-significant correlation when {$[R,~\sigma]$~=~[200~Mpc,~50~Mpc]}. $r_s$ is increased using real-space galaxy positions. As would be expected, the real-space and redshift-space results {differ} most when considering the density on small scales.

We test for the effects of the {$0.02 < z < 0.05$} SN selection by instead including {$z < 0.02$} SNe. We also alleviate the galaxy luminosity cut to {$L_{z} > 7.0$}, to test for the effects of increasing the number of tracers of the density field. These results are shown in yellow and magenta, respectively. In both cases, no significant change to the amplitude of $r_s$ as a function of $\Delta \phi_{+-}$ is found. {In the case of including {z < 0.02} SNe, this implies that although we are forced to omit these lowest-redshift SNe in the observations due to uncertainties in peculiar velocity, they are not crucial for an assessment of $\Delta \phi_{+-}$.} In the case of the increased number of tracers, this implies that the density field is already sufficiently sampled for {$L_{z} > 8.843$}, and hence, so too is the 2MRS sample.

We saw that using a finite value of $\sigma$ increased values of $r_s$ for large sphere radii, $R$. As such, we test the effects of fixing $R = 200$ Mpc, and instead vary $\sigma$ between 10 and 200 Mpc. The result, shown in green in Figure \ref{fig:phi_rs}, reveals that a density weighting corresponding to {$\sigma \sim 40$ Mpc} produces the maximum significance of correlation between {fractional $H_{0}$ error} and $\Delta \phi_{+-}$. {Note that we make qualitatively identical conclusions to those found in Figure \ref{fig:phi_rs} when plotting the {$p$-value} associated with a correlation against $R$ and $\sigma$}. 

The underlying result of these analyses is that density gradients on super-cluster scales {$\sim 50$ Mpc} are most strongly correlated with estimates of {fractional $H_{0}$ error}. This result is in concordance with expectations from the well-known $J_{3}(r)$ integral \citep[see, e.g.][]{PEE81}. The 2-point correlation function of galaxies together with linear theory predicts that the largest contribution to peculiar velocities comes from density structures on these scales \citep{CBP81}. It is also noted that this scale size is established to maximise angular diameter distance biases via gravitational deflection \citep{KP16}, which is albeit a small gravitational lensing effect. These factors support conclusions that the correlations between density structure on supercluster scales and $H_{0}$ are in fact due to real gravitational effects.

We note that a sphere size of {$R = 200$ Mpc} is not appropriate for the case of the observations, as a large fraction of the {$0.02 < z < 0.05$} Pantheon SNe lie within 200 Mpc of the ZoA. In the observations, as spheres around SNe which overlap the survey edge may produce unreliable $\Delta \phi_{+-}$ measurements, one may expect that this is why the prescription {$[R, \sigma]$ = [50 Mpc, 50 Mpc]} was instead found to be optimal. {We reiterate, however,} that in the $z$-space simulations, $r_s$ flattens out for {$R > 50$ Mpc}, suggesting that the trend of {fractional $H_{0}$ error} with $\Delta \phi_{+-}$ would not improve significantly in the observations were we able to access a greater volume. As a result, we expect that we have found close to the maximum coherence of {fractional $H_{0}$ error} with $\Delta \phi_{+-}$ with the {$[R, \sigma]$ = [50 Mpc, 50 Mpc]} prescription.

For the next stage of our analysis{,} we again use the mock redshift-space galaxy density field, {$[R, \sigma]$ = [50 Mpc, 50 Mpc]}, and for 100 random observer positions in the box, we each time draw \newthree{88 mock SNe from the simulation, in order to match to the number of Pantheon SNe which are observed at {0.02 < $z$ < 0.05} and at least 50 Mpc from the ZoA}. This enables us to test for the effects of sample size on our $H_{0}$ vs $\Delta \phi_{+-}$ correlation.

\begin{figure}
    \centerline{\includegraphics[width=1.\columnwidth]{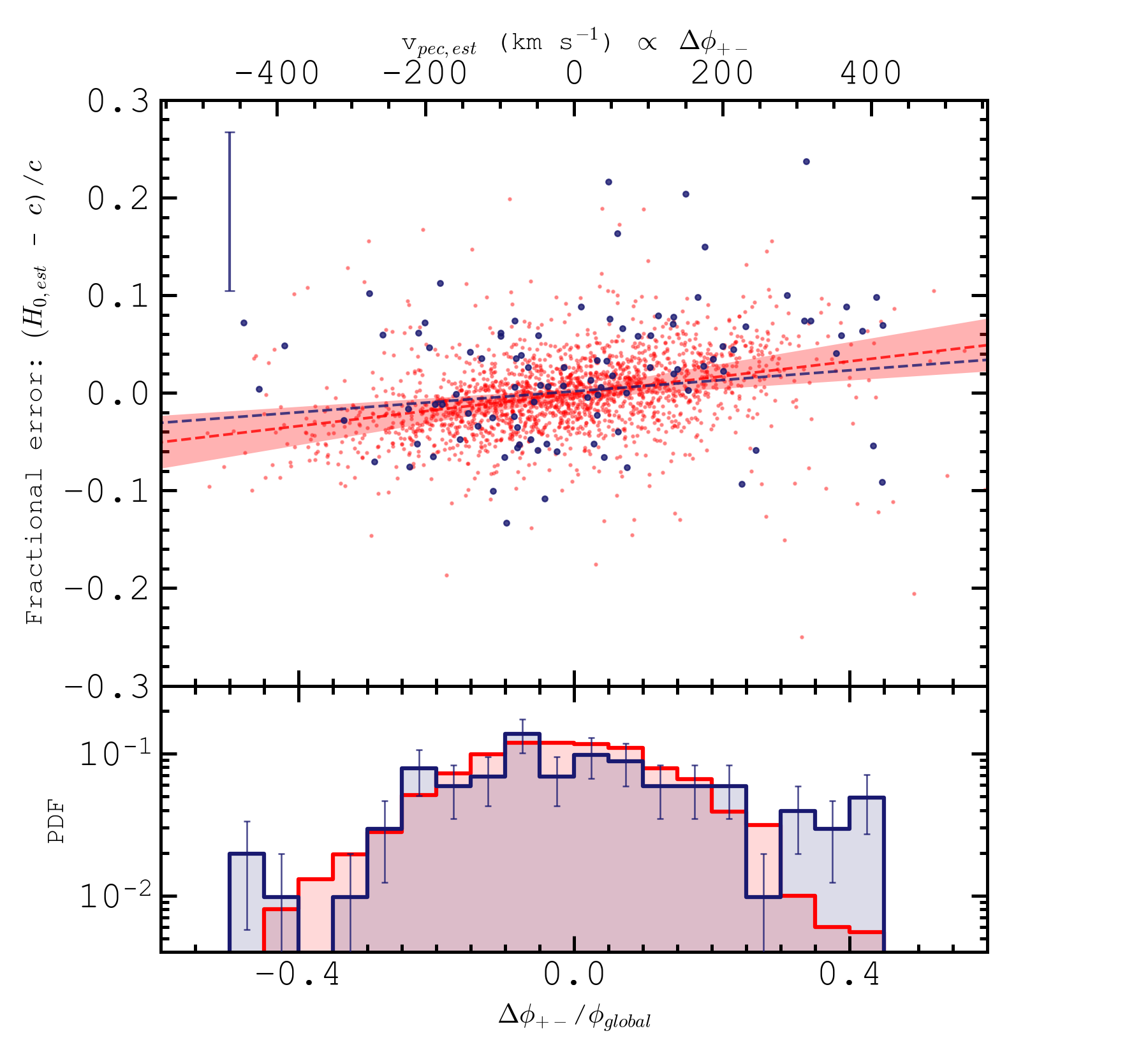}}
    \caption{Top: {fractional $H_{0}$ errors} for {{0.02 < $z$ < 0.05}} SNe as a function of {$\Delta \phi_{+-}$}, using {[$\sigma$,$R$]=[50 Mpc, 50 Mpc]} (see text \& Figure~\ref{fig:7}). Blue points show \newthree{88} observed Pantheon SNe, where the median uncertainty on {fractional $H_{0}$ error} is shown as the blue error bar at the top-left of the panel. Red points represent 2000 mock SNe from the MDPL2-Galacticus model, each viewed from a random observer position. The blue dashed line shows the error-weighted line of best fit to the observational data. The red dashed line and filled region depict the mean and standard deviation in the best-fit line, respectively, to the mock data when matching the observational sample size of {{$N_{sn}$ = \newthree{88}}}, averaged over 100 Monte Carlo iterations and observer positions. The secondary x-axis (top) shows estimates of $v_{pec}$ as a function of ($\Delta \phi_{+-}$), inferred from the gradient of the linear fit of $v_{pec}$ to $\Delta \phi_{+-}$ in the mock data. Bottom: \newthree{Probability distributions of {$\Delta \phi_{+-}$} for 2000 simulated SNe (red) and 88 Pantheon SNe (blue). Poisson errors on the observed result are shown as blue error bars.}}
    \label{fig:8}
\end{figure}

For each iteration, a linear fit of {fractional $H_{0}$ error} to $\Delta \phi_{+-}$ is taken. Figure~\ref{fig:8} shows with red dashed line the mean gradient and intercept values, averaged over the 100 iterations. {The intercept is allowed to vary for each iteration, but the mean intercept over iterations is set to 0 at {$\Delta \phi_{+-}$ = 0}}. The red filled region shows the standard deviation in the regression line parameters over the iterations. The {\newthree{88} {$0.02 < z < 0.05$} Pantheon SNe} are shown as blue points, and the blue dashed line depicts the regression line to the data seen in Figure~\ref{fig:7}. The observational and simulated results show excellent consistency for the slope of local {fractional $H_{0}$ error} with $\Delta \phi_{+-}$. {Note that we also assumed a 737 cosmology when calculating fractional $H_0$ errors in the models. However, once again the results are not sensitive to the fiducial $H_0$ assumed.}

The mean slope for the simulations, of {$\overline{\mathcal{S}} = 0.061 \pm 0.021$}, implies with 3$\sigma$ confidence that the observer will find a positive trend of $H_{0}$ estimates with $\Delta \phi_{+-}$ at a random observer position in the Universe {when using a SN sample of matching statistics to the Pantheon sample}. {This is consistent with the observation slope of $\overline{\mathcal{S}} = 0.065$.} {Although separate from the analysis of fractional $H_0$ offset due to peculiar velocity}, note that the mean intercept in the simulations is found to be {$\overline{c}~=~69.99$~km~s$^{-1}$~Mpc}, with a root mean square (rms) deviation from the fiducial {$H_{0}~=~70$~km~s$^{-1}$~Mpc$^{-1}$} of {{0.26}~km~s$^{-1}$~Mpc$^{-1}$}, showing that a regression fit reproduces the fiducial $H_{0}$ at {$\Delta \phi_{+-}$~=~0}, {and hence, when there is zero peculiar velocity}. The rms error from the model is an estimate of the cosmic variance in the trend of $H_{0}$ estimates with $\Delta \phi_{+-}$. The mean values of $r_s$ and $p$ are {\newthree{0.4010} and \newthree{0.0006}}, respectively. 

In the simulations we are free from {uncertainties from SN photometry and from light-curve fitting techniques}, which result in the larger spread in observational {fractional $H_{0}$ errors} compared with results from the model. This highlights the fact that uncertainty in the SN photometry is what limits the significance of our observed correlation to \newthree{$r_{s} = 0.2739$}, rather than the ability of $\Delta \phi_{+-}$ to capture peculiar velocity information. 

Recalling that the mock sample used for these calculations is luminosity limited, we repeat tests for the trend of {fractional $H_{0}$ error} vs $\Delta \phi_{+-}$, but with a flux-limit and {corresponding} galaxy weighting procedure employed, as seen in Section \ref{subsec:regionalrho}, to test for the effects of galaxy weighting on our observational correlations. We choose the mock flux limit to be at a magnitude of $m_z = 15.89$, such that the galaxy sample starts to become incomplete at a redshift $z = 0.0202$, as found for the observations. We find that there is no significant change to the slope of {fractional $H_{0}$ error} vs $\Delta \phi_{+-}$ when using a mock flux limit, nor does the {cosmic} variance on the intercept increase. This implies that the weighting of galaxy statistics as a function of redshift, required for our observational density calculations, has a negligible effect on the magnitude and uncertainty of $H_{0}$ estimate corrections.

Reverting to the luminosity-limited sample, we also show in Figure~\ref{fig:8}, as the red points, $\Delta \phi_{+-}$ vs {fractional $H_{0}$ error} for 2000 simulated SNe. Here, the observer's position is changed for each observation. These data follow tightly the mean regression line found for the mock data using \newthree{$N_{sn} = 88$}. The bottom panel shows \newthree{the probability distribution} of the 2000 $\Delta \phi_{+-}$ values \newthree{in red}, showing that the mean $\Delta \phi_{+-}$ value over all observer positions is close to zero. \newthree{The blue probability function shows that the distribution of $\Delta \phi_{+-}$ values from Pantheon SNe is consistent with the model distribution, within the Poisson errors shown.}

We can use our knowledge of SNIa peculiar velocities in the mock data to relate this velocity to its proxy, $\Delta \phi_{+-}$. For the 2000 randomly selected SNe, we find that the regression line $v_{pec} = 618.5 \Delta \phi_{+-}$ best approximates the relation. Using this scaling, we plot an estimate of peculiar velocity as a secondary x-axis in the top panel of Figure~\ref{fig:8}. Our scaling, coupled with the $\Delta \phi_{+-}$ distribution shown in the bottom panel of Figure~\ref{fig:8} implies that the $1\sigma$ deviation from zero peculiar velocity is {$\sim$~120~km~s$^{-1}$}, i.e. 68\% of SN positions are estimated to have an absolute peculiar velocity less than this value. {From this scaling}, the observational SN positions are estimated to have a mean absolute peculiar velocity of {$\sim$~100~km~s$^{-1}$}, with a standard deviation of {$\sim$~75~km~s$^{-1}$}.

In conclusion we have found, using the MDPL2-Galacticus simulation, reassuring consistency for the trend of {fractional $H_{0}$ error} estimates vs $\Delta \phi_{+-}$ when compared with the observational results from the {Pantheon SN sample} and 2MRS galaxies. We have used these simulations to compute the expected cosmic variance in the trend of {fractional $H_{0}$ error} with $\Delta \phi_{+-}$, to inform us of the expected uncertainty on any $H_0$ estimates when corrected for density effects.

{The error-weighted mean value of fractional $H_0$ error for the \newthree{88} Pantheon SNe is found to be \newthree{$6 \times 10^{-4}$}. Given that the fractional error is defined to be zero at $\phi_{+-} = 0$, this means that in the case of this SN sample, peculiar velocities affect the mean estimate of $H_0$ by $< 0.1 \%$. This result shows that with a large number of SNe and sufficient sky coverage, the net effect of peculiar velocities on the mean $H_0$ estimate from SNe is negligible.}

{\subsection{Calibration of SNIa distance moduli \& an estimate of $H_0$}\label{sec:BAO}}

{The main focus of this paper has been the fractional effect on $H_0$ measurements from peculiar velocities. However, for completeness, we estimate an $H_0$ value from our {$0.02 < z < 0.05$} sample of Pantheon SNe.}

{To estimate $H_0$ with Equation~\ref{eq:ivanH0}, we rely on the accuracy of our SN distance moduli. To calibrate the distance moduli, we utilise the {$z = 0.57$} angular diameter distance ($D_A$) result of \citet{AND14}, derived from detections of baryon acoustic oscillations (BAO) in the clustering of galaxies. \newthree{$D_A$ can be represented as $1421 \pm 20$ Mpc ($r_d/r_{d,fid}$) where {$r_{d,fid} = 149.28$ Mpc} is the fiducial sound horizon scale used by \citet{AND14}. This can be converted to an equivalent distance modulus using {$D_L = D_A (1+z)^2$}, leading to $\mu = (42.72 \pm 0.03) + 5 \log (r_d/r_{d,fid})$ mag.}}

{We next turn to a higher redshift portion of the Pantheon sample, in order to have a sample covering the redshift of the BAO result. To avoid an assumption for $M_{B,fid}$ (the fiducial stretch and colour corrected SNIa absolute magnitude) which is degenerate with $H_0$, we perform a linear fit of the corrected apparent magnitude ($\mu_{B} + M_{B,fid}$) against the logarithm of CMB-frame redshift, for 118 SNe in the redshift range {0.45 < $z$ < 0.70}. We then determine the offset to {$\mu_{B} + M_{B,fid}$} required for this fit to intercept the BAO-derived distance modulus at {$z = 0.57$}. We find that \newthree{$\mu_{B} + M_{B,fid}$ + $19.45 \pm 0.04 + 5 \log (r_d/r_{d,fid})$} coincides with the BAO result, and so correct the lower redshift SN distance moduli accordingly. This calibrates the SNe distance moduli using the BAO scale with negligible dependence on cosmology or peculiar velocities since we interpolate to the {$z=0.57$} BAO result using only data from {0.45 < $z$ < 0.70}.}

{Returning to the now calibrated ${0.02 < z < 0.05}$ sub-sample, from Equation~\ref{eq:ivanH0}, the set of $H_0$ estimates uncorrected for peculiar velocities can be found. The error-weighted mean value of $H_0$ before peculiar velocity correction is \newthree{$H_{0}~=~(67.47~\pm~1.00) \times (r_{d,fid}/r_{d})$~km~s$^{-1}$~Mpc$^{-1}$}.}

{In Section~\ref{sec:H0U21} we estimated the observational slope, $\mathcal{S}$, of fractional $H_{0}$ error vs $\Delta \phi_{+-}$, and in Section~\ref{sec:Sims}, the uncertainty in this result due to cosmic variance, given our SN sample size. Converting $\mathcal{S}$ to units of {km~s$^{-1}$~Mpc$^{-1}$}, we can calculate individual peculiar-velocity corrected values as $H_{0,corr} = H_{0} - \mathcal{S} \Delta \phi_{+-}$. The error-weighted mean $H_{0}$ measurement over the SN sample is our best estimate for the present-day value of the Hubble parameter.}

{We utilise a $10^{4}$ iteration MC technique to compute our {best-estimate and its uncertainty}. We vary the density-corrected SN $H_{0}$ measurements for each iteration given uncertainties in the slope, $\mathcal{S}$, estimated from the simulations. We also fold in uncertainties in the SN photometry and in the re-calibration of SN distance moduli {to the BAO-inferred distance scale}. {We calculate the error-weighted mean of the \newthree{88} individual $H_{0}$ estimates for each iteration. Our best estimate is then given by the mean and standard deviation of this average over the iterations.}}

\begin{figure}
    \centerline{\includegraphics[width=1.\columnwidth]{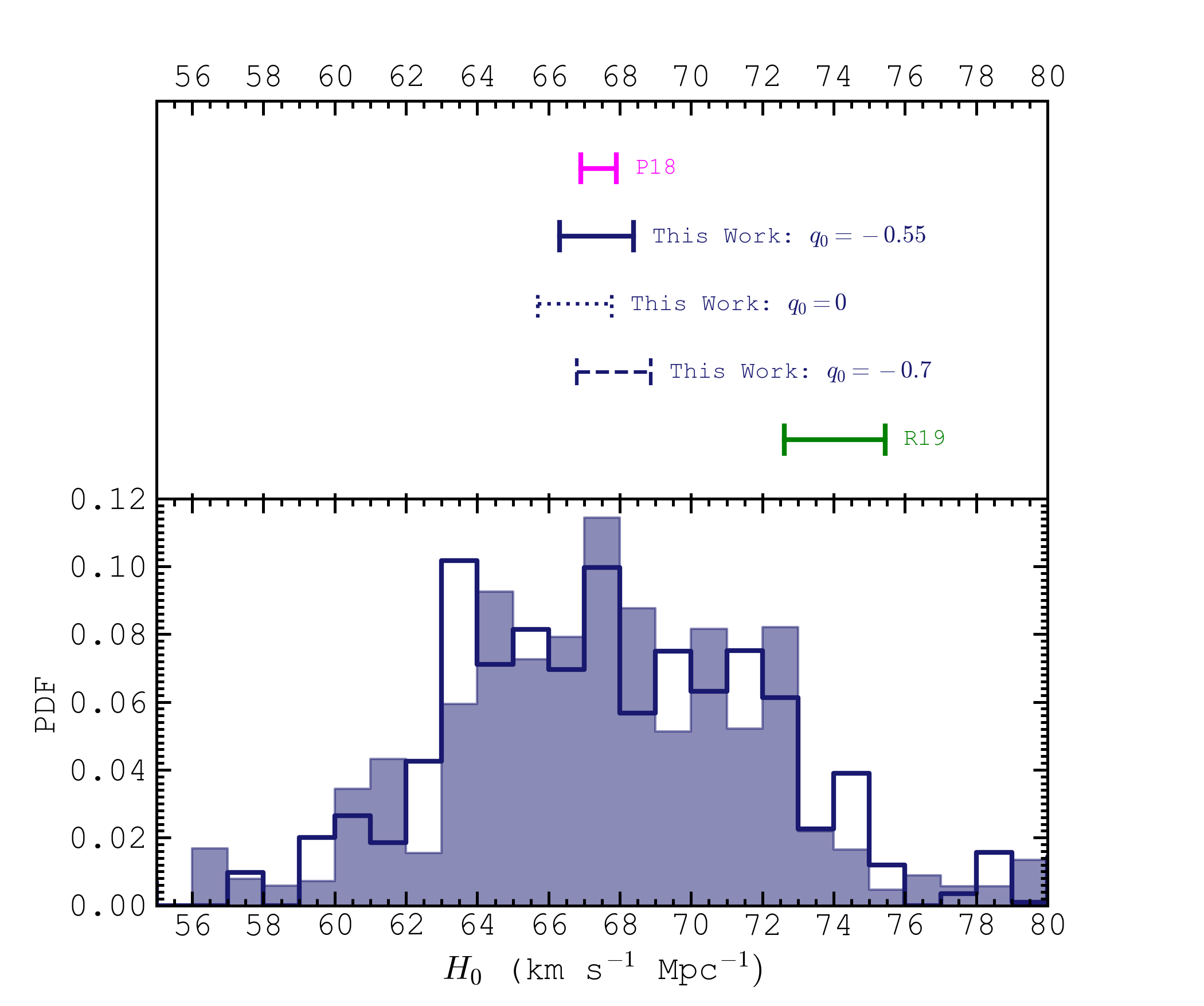}}
    \caption{{Top: a comparison of $H_{0}$ estimates. In blue: the results of the present work derived from SNe with distance moduli {calibrated on the BAO-inferred cosmic distance scale (see text)}, and corrected for peculiar velocity effects. The blue solid range indicates the 1$\sigma$ uncertainty on our best $H_{0}$ estimate from \newthree{88} Pantheon SNe, where errors account for uncertainties in SN photometry, in SN distance calibration, and in the cosmic variance of peculiar velocity effects. This result assumes $q_{0} = - 0.55$. For comparison, the dotted and dashed ranges depict the change to this result, assuming instead $q_{0}=0$ and $q_{0}=-0.7$, respectively.} In green: the R19 $H_{0}$ measurement from a combination of LMC DEBs, masers in NGC 4258 and Milky Way parallaxes. In pink: the P18 result from the CMB and {$\Lambda$-CDM}. {Bottom: the unfilled (filled) histogram represents the error-weighted probability density function (PDF) of individual $H_{0}$ estimates from the SNe using Equation \ref{eq:ivanH0}, before (after) corrections for the effects of peculiar velocities.}}
    \label{fig:9}
\end{figure}

{We infer that \newthree{$H_{0}~=~(67.41~\pm~1.02)~\times~(r_{d,fid}/r_{d})$~km~s$^{-1}$~Mpc$^{-1}$}, as shown by the solid blue range in Figure~\ref{fig:9}. This result is consistent with that obtained by \citet{P18}, who find {$H_{0}~=~67.40~\pm~0.50$~km~s$^{-1}$~Mpc$^{-1}$}. Conversely, our result lies in 3.8 $\sigma$ tension with the result of \citet{R19}, who find {$H_{0}~=~74.03~\pm~1.42$~km~s$^{-1}$~Mpc$^{-1}$}, using LMC Cepheid standards to calibrate the distance scale and constrain distance moduli of SNeIa residing in Cepheid hosts. {We conclude that the Pantheon SN sample} is large enough and surveys a large enough volume that the sign of peculiar velocities is unbiased, and therefore that accounting for estimated peculiar velocities of Pantheon SNe does not resolve the Hubble tension.}

{The corrected $H_0$ distribution for the individual SNe is also shown, as the filled histogram. It is once again clear from comparison with the uncorrected distribution that the net effect of peculiar velocities on the average $H_0$ estimate is small when averaged over a large number of SNe at different sky positions, with a negligible reduction to the mean $H_0$ value of only \newthree{$0.06~\times~(r_{d,fid}/r_{d})$ mag} via this correction.}

{The component of the error in our $H_{0}$ estimate due to peculiar velocity corrections has a magnitude of \newthree{$0.26~\times~(r_{d,fid}/r_{d})$~km~s$^{-1}$~Mpc$^{-1}$} when accounting for the model-estimated cosmic variance in the slope of $H_{0}$~vs~$\Delta \phi_{+-}$. The vast majority of this error is found to stem from noise in the $v_{pec}$ vs $\Delta \phi_{+-}$ relation, which introduces error in the $H_{0}$ estimate vs $\Delta \phi_{+-}$ relation. Variations in the $H_{0}$ distribution over observer positions are found to have a relatively negligible contribution to the error.}

{The resultant error in our best $H_{0}$ estimate using Pantheon SNe is the quadrature sum of: i) an error of \newthree{$0.95~\times~(r_{d,fid}/r_{d})$~km~s$^{-1}$~Mpc$^{-1}$ from BAO-based calibrations of SN distance moduli}; ii) an error of \newthree{$0.33~\times~(r_{d,fid}/r_{d})$~km~s$^{-1}$~Mpc$^{-1}$} from SN photometric uncertainties; and iii) an error of \newthree{$0.26~\times~(r_{d,fid}/r_{d})$~km~s$^{-1}$~Mpc$^{-1}$} from our corrections of the $H_{0}$ estimates for peculiar velocity effects. Thus, for comparable SN samples and for future samples with larger statistics and coverage, density effects are not expected to be the main cause of the Hubble tension. {Instead, the majority of the uncertainty on the local $H_{0}$ estimate stems from uncertainties in the calibration of SNIa photometry.}}

All the results discussed have adopted $[q_{0},~\Omega_{m},~\Omega_{\Lambda}]~=~[-0.55,~0.7,~0.3]$. For comparison, using instead [$q_{0},~\Omega_{m},~\Omega_{\Lambda}]~=~[0,~0.67,~0.33]$ causes a \newthree{$0.68~\times~(r_{d,fid}/r_{d})$~km~s$^{-1}$~Mpc$^{-1}$} drop in our best-estimate $H_{0}$ to {66.73~km~s$^{-1}$~Mpc$^{-1}$}. Adopting [$q_{0},~\Omega_{m},~\Omega_{\Lambda}]~=~[-0.7,~0.2,~0.8]$ causes a \newthree{$0.42~\times~(r_{d,fid}/r_{d})$~km~s$^{-1}$~Mpc$^{-1}$} rise, giving \newthree{$H_{0}~=~67.83~\times~(r_{d,fid}/r_{d})$~km~s$^{-1}$~Mpc$^{-1}$}. These results are shown as the blue dotted and dashed ranges in the top panel of Figure~\ref{fig:9}, respectively, {and demonstrate that errors from SN distance calibration dominate the error budget as opposed to errors associated with the fiducial cosmology at these low redshifts.}

{Finally, we emphasise that even though we calibrated the SNe to the inverse distance ladder, the same relative effects on the result due to peculiar velocities would be evident were the SNe calibrated to the local distance ladder.}
\\
\\
\section{Summary \& Conclusions}

Using the ${K < 11.75}$ flux-limited 2MASS redshift survey (2MRS) of galaxies \citep{HUC12}, and assuming that the $K$-band luminosity distribution is well-approximated by a Schechter function, we use the STY maximum-likelihood method \citep{STY79} to infer a best-fit Schechter function to the data with parameters {[$\alpha$, $L^{*}$] = [--0.99, 10.97]}, fitting the data well as a function of redshift when accounting for galaxy luminosity evolution effects. This yields {$L_{K} > 10.5$} sample completeness as a function of redshift, allowing a reconstruction of the galaxy density field. Whilst we find region-specific density structure which is qualitatively consistent with the findings of WS14 and \citet{BOH19}, we find no strong evidence for a `Local~Void' which pertains to the whole sky, {out to the $z = 0.1$ redshift limit of the 2MRS galaxy survey}, in agreement with \citet{CAR15}.

We have introduced a density parameter, denoted here as $\Delta \phi_{+-}$, which quantifies density gradients along a LOS. $\Delta \phi_{+-}$ is a proxy for peculiar velocities as a function of location in the local Universe. Using a sample of {\newthree{88} SNeIa from the Pantheon sample \citep{SCO18}}, in a redshift range {{$0.02~<~z~<~0.05$}}, we see the clear effects of the density field on $H_{0}$ estimates, from trends of fractional $H_{0}$ error vs $\Delta \phi_{+-}$. We find from this empirical method that density gradients on the scale of super-clusters ($\sim$~50~Mpc) have the strongest effects on {local fractional $H_{0}$ errors}.

We use the present-day snapshot from the MDPL2-Galacticus Simulation \citep{KNE18} to repeat our analysis with a mock galaxy density field and SN sample, which is free from photometric uncertainties, and find remarkably consistent results with the observations for the trend of {fractional $H_{0}$ errors with $\Delta \phi_{+-}$}. Maximum coherence between {fractional $H_{0}$ error} and $\Delta \phi_{+-}$ is again found for density structure on the scale of super-clusters ($\sim$~50~Mpc), coincident with expectations from the behaviour of the correlation function of galaxies \citep[see, e.g.][]{CBP81}, increasing confidence that these strong correlations are in fact due to real gravitational effects.

We find that the {{$0.02~<~z~<~0.05$} Pantheon sample has enough SN statistics and survey volume that the mean peculiar velocity of these SNe lies close to zero}. As a consequence, the average offset in $H_{0}$ estimates due to galaxy density effects is also close to zero. We use the simulations to estimate the cosmic variance in the peculiar velocity distribution when matching to the sample size and sky coverage of the observations, finding that the mean peculiar velocity for such a sky coverage and sample size lies close to zero over practically all observer positions. However, {should one wish to estimate $H_{0}$ using local SN surveys which are not all sky}, we note that our method would be able to correct for the effects of the density field on $H_{0}$ estimates, irrespective of peculiar velocity biases.

In terms of the methods of the present work, analyses of biases in fractional $H_{0}$ error estimates can be built upon with various improvements to assessments of the galaxy density field. These improvements could include: a replacement of 2MRS with 2M++ galaxies \citep{LH11}; an assessment of the density structure within the `Zone of Avoidance' \citep{HUB34}; and increased magnitude-depth of all-sky near-IR galaxy surveys from, for example, the UKIRT Hemisphere Survey \citep{DYE18}, the VISTA Hemisphere Survey \citep{SUT15} and LSST \citep{IVE19}. Assessments of galaxy cluster densities from deep X-ray surveys such as eROSITA \citep{MER12} also promise to put state-of-the-art constraints on the local density structure. With the ability to probe the density field over a larger redshift range one can also examine evidence for voids out to cosmological distances for tens of thousands of galaxies or clusters, as well as the relationship of any structure with standard-candle $H_{0}$ estimates. Note that as advances in photometric precision and distance calibration techniques arrive, studies of the effects of the density field and resultant peculiar velocities will become increasingly important for local measurements of the Hubble constant.

\section{Acknowledgments}

TMS acknowledges support from an STFC DTP studentship, jointly supported by the Faculty of Engineering and Technology at LJMU.
CAC acknowledges support from LJMU and STFC for resources to conduct the research described here.

This publication has made use of the following resources:

\begin{enumerate}
    \item The Two Micron All Sky Survey, which is a joint project of the University of Massachusetts and the Infrared Processing and Analysis Center at the California Institute of Technology, funded by the National Aeronautics and Space Administration and the National Science Foundation.

    \item The CosmoSim database, a service by the Leibniz-Institute for Astrophysics Potsdam (AIP). The MultiDark database was developed in cooperation with the Spanish MultiDark Consolider Project CSD2009-00064. The authors gratefully acknowledge the Gauss Centre for Supercomputing e.V. (\url{www.gauss-centre.eu}) and the Partnership for Advanced Supercomputing in Europe (PRACE, \url{www.prace-ri.eu}) for funding the MultiDark simulation project by providing computing time on the GCS Supercomputer SuperMUC at Leibniz Supercomputing Centre (LRZ, \url{www.lrz.de}). The Bolshoi simulations have been performed within the Bolshoi project of the University of California High-Performance AstroComputing Center (UC-HiPACC) and were run at the NASA Ames Research Center.
\end{enumerate}

\section{Data Availability}
2MRS data was obtained at \url{tdc-www.harvard.edu/2mrs}.
MDPL2-Galacticus data was obtained via SQL query at \url{www.cosmosim.org}.
Pantheon SN data was obtained at \url{archive.stsci.edu/prepds/ps1cosmo}.
Data products of this article will be shared on reasonable request to TMS.

\bibliography{main.bib}

\appendix

\section{A second-order Hubble Law}\label{sec:Ivancontd}

The Hubble law is often stated such that the recession velocity is equal to the Hubble constant times the distance, 
with the most common approximation for velocity given by $c z$.
However, a more useful expression for velocity \citep[e.g.][]{CAP17,EMS18} is given by 
\begin{equation}
 v = c \ln(1+z) \: .
\end{equation}
This is more accurate for pure line-of-sight velocity 
and means that the peculiar velocity and cosmological terms, 
and frame corrections, are additive \citep{BAL18}. 
A common misconception is to assume $cz$ terms are additive. 
Coupled with different distance definitions, 
there are thus many versions of a Hubble law. 

\begin{figure}
\centerline{\includegraphics[width=1.\columnwidth]{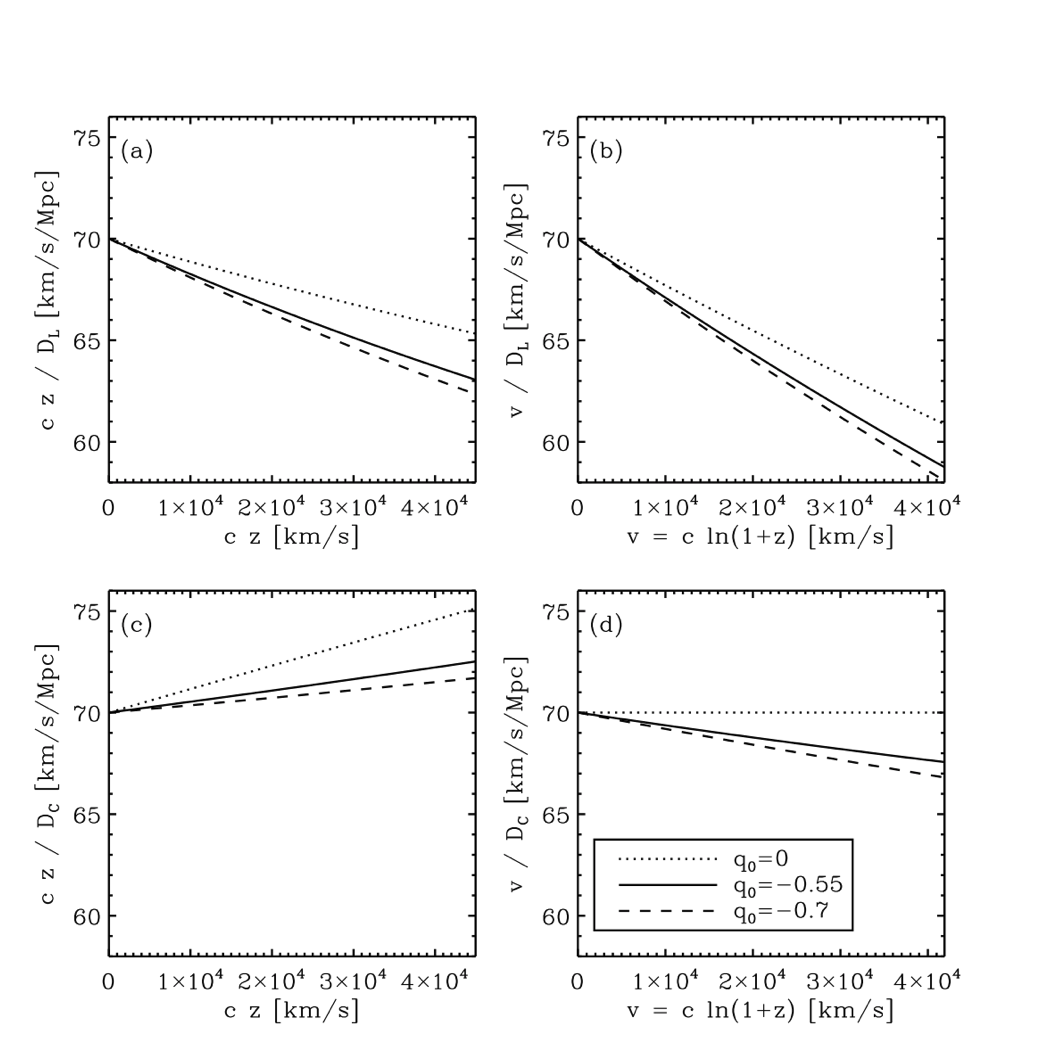}}
\caption{Different views of the Hubble law.
  The relations shown are for: 
  a coasting cosmological model ($q_0=0$), 
  a flat $\Omega_{m,0}=0.3$ model ($q_0=-0.55$), and 
  a flat $\Omega_{m,0}=0.2$ model ($q_0=-0.7$).}
\label{fig:hlaws}
\end{figure}

Figure~\ref{fig:hlaws} shows four different views of the
Hubble law using these approximations for velocity 
with luminosity distance ($D_L$) and line-of-sight comoving distance ($D_C$). 
For each version, curves are shown for three model cosmologies, 
all with flat geometry and with {$H_{0}=70$~km~s$^{-1}$~Mpc$^{-1}$}.
Two are $\Lambda$CDM models, 
for which the deceleration parameter $q = {\Omega_m}/{2} - \Omega_{\Lambda}$,
while the other is a `coasting' model with $w=-1/3$. 
Notably, none of these versions of the Hubble law are accurate 
except in the case of (d) $v = c \ln(1+z) = H_{0} \, D_C$ for
the coasting model \citep{SR15}.
Note this exact law also is valid for a non-flat coasting model such as an empty universe
[though in this case, $D_L \neq (1+z) D_C$]. 
Below we show a derivation of a second-order Hubble law that 
is natural in this view with a transparent dependence on $q_0$.

For demonstration purposes, we consider a flat universe with a single type of fluid 
with equation of state $w$ such that: 
\begin{equation}
 q \:=\: \frac{1+3w}{2}  \mbox{~~~and~~~}  E(z) \:=\: (1+z)^{q+1}  \mbox{~~~.}
\end{equation}
The comoving distance is then given by 
\begin{equation}
\comovingdist 
  \:=\: \frac{c}{H_0} \int_0^z \frac{\dd z}{E(z)}
  \:=\: \frac{c}{H_0} \int_0^z \frac{\dd z}{(1+z)^{q+1}}  \mbox{~~~.}
\end{equation}
Using the logarithmic shift $\zeta = \ln (1+z)$, this becomes 
\begin{equation}
\comovingdist 
  \:=\: \frac{c}{H_0} \int_0^\zeta \frac{(1+z)}{E(z)}  \dd \zeta
  \:=\: \frac{c}{H_0} \int_0^\zeta e^{-q \zeta} \, \dd \zeta  \mbox{~~~;}
\end{equation}
and after integrating ($q \neq 0$), 
\begin{equation}
\comovingdist \:=\: \frac{c}{H_0} \left[ \frac{1}{q} (1 - e^{-q \zeta} ) \right]   \mbox{~~~.}
\label{eqn:cdist-w}
\end{equation}

For a non-constant $q$, the above result is valid only over a small change in $\zeta$. 
For small $\zeta = v/c$, using a second-order Taylor series expansion, we obtain 
a second-order Hubble law: 
\begin{equation}
\comovingdist 
  \:\simeq\: \frac{c}{H_0} \, \zeta \, \left( 1 - \frac{q_0 \zeta}{2} \right)
  \:=\:      \frac{v}{H_0}             \left( 1 - \frac{q_0 v}{2 c} \right) \mbox{~~~.}
  \label{eqn:second-order-h-law}
\end{equation}
This form tends to the exact law with $q_0 \rightarrow 0$, 
and the right-hand term [$1 - (q_0/2)(v/c)$] represents an average 
of $(1+z)/E(z)$ assuming constant acceleration 
(c.f.\ the quadratic fitting function given by \citet{SR15} for improved precision).

For $\Lambda$CDM cosmologies, the approximation is accurate to within 0.1\% at $z \lesssim 0.1$. 
Note that regardless of the accuracy of the Hubble law, 
$v$ accurately represents the integral of the velocity differences along the line-of-sight, 
precisely in the case of fundamental observers. This is evident from the additive nature of 
terms in $\zeta$ or $v$ \citep{BAL18}. 

\section{{Studying the SGC local underdensity via simulations}}\label{sec:SGCsims}

{In Section~\ref{subsec:regionalrho}, we concluded that we find no evidence for a `Local~Void' which pertains to the full sky {out to the $z = 0.1$ limit of the 2MRS survey}. However, we found a significant under-density in the direction of the SGC-6dFGS region, which was $27 \pm 2 \%$ under-dense integrated below $z < 0.05$.}

{In Section~\ref{sec:Sims} we demonstrated a useful property of the MDPL2-Galacticus simulations: we were able to estimate the cosmic variance of peculiar velocity effects on $H_{0}$ estimates by mimicking our observational analysis from variety of mock observer positions. Using a similar method, we can test for how `common' the SGC-6dFGS under-density is, by testing how often an under-density of this amplitude is observed at different observer positions in the simulation. We place the observer at $10^6$ random positions in the 1 $h^{-3}$ Gpc$^{3}$ box.}

\begin{figure}
    \centerline{\includegraphics[width=1.0\columnwidth]{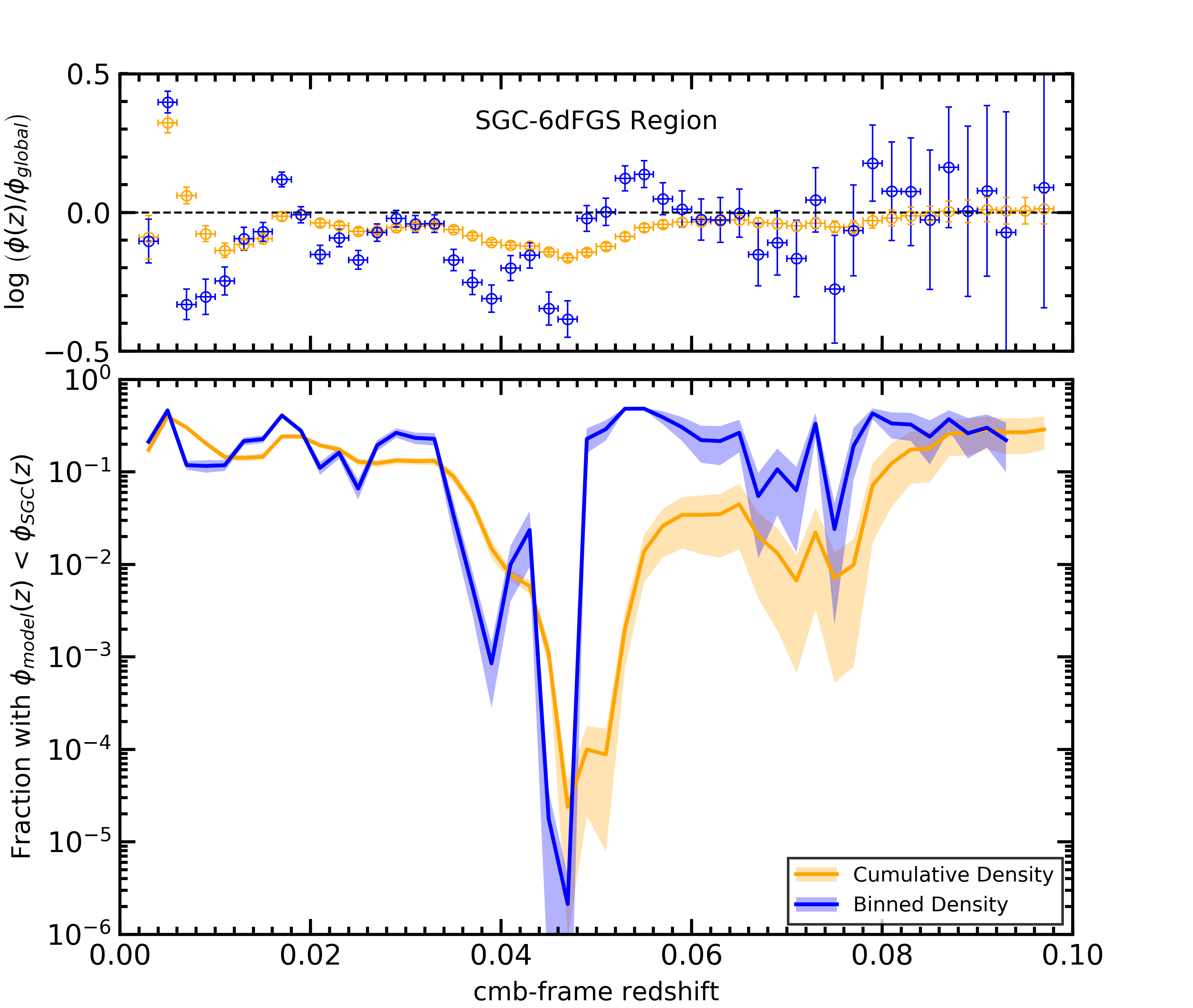}}
    \caption{{Top: local galaxy densities as a function of redshift in the SGC-6dFGS region, in logarithmic units of the global density. In blue is the binned density at $z$. In yellow is the cumulative (integrated) density out to $z$. Redshift bins are of width 0.002. Bottom: the fraction of $10^6$ SGC-6dFGS-sized regions from the MDPL2-Galacticus simulation, which have a density less than that observed in SGC-6dFGS, as a function of redshift. The simulated densities use a mock detection limit matching the observations, and observer position is randomised for each iteration.}}
    \label{fig:lessSGC}
\end{figure}

{Figure~\ref{fig:lessSGC} re-iterates that the main contribution to SGC-6dFGS under-density occurs at $z \sim 0.05$. For integrated densities out to $z \sim 0.03$, the percentage of mock observed positions which produce a density at least as under-dense as found in SGC-6dFGS is approximately $40\%$. This implies such an under-density is common-place at most positions in the universe at current-epochs. For clarity, were we to compare the simulated densities to the global density, we would find this percentage stays close to $50\%$ across the redshift range, as the density at a given redshift is equally likely to be over-dense as under-dense at a random observer position.}

{However, what is striking about the results of Figure~\ref{fig:lessSGC} is that the SGC-6dFGS under-density at $z \sim 0.05$ is extremely unlikely to arise from the vast majority of mock observer positions: The number of the $10^6$ positions finding such an integrated underdensity out to $z = 0.05$ is of the order 100, or 0.01\%. This either implies that our position in the Universe is particularly special, that the large scale structure in the simulation is unrealistic, or that there is an unknown observational systematic in the direction of SGC-6dFGS. Given that several more studies, including \citet{WS14}, find the $z = 0.05$ SGC-6dFGS under-density to be of notably high-amplitude, a quantification of such a systematic in future work would be of great interest. {However, we note that the correspondence with the X-ray REFLEX clusters result argues against it being a systematic associated with the galaxy surveys.} The tension between measurements of the local under-density and the current cosmological model highlights the great potential in future work using deeper and more complete extra-galactic samples with new facilities such as eROSITA.}

\label{lastpage}
\end{document}